\documentclass[prb,aps,amsmath,amssymb,amsfonts,floatfix,superscriptaddress,showpacs,twocolumn,nofootinbib]{revtex4-1}
\usepackage{graphicx}
\usepackage{color}
\usepackage[utf8]{inputenc}
\usepackage{hyperref}

\newcommand{\av}[1]{\ensuremath{\left\langle #1 \right\rangle}}
\newcommand{\up}{\ensuremath{\uparrow}}
\newcommand{\dn}{\ensuremath{\downarrow}}

\renewcommand{\Im}{\operatorname{Im}}

\begin{document}
 
\title{Capturing non-local interaction effects in the Hubbard model:\\ optimal mappings and limits of applicability}

\author{E. G. C. P. van Loon}
\affiliation{Radboud University, Institute for Molecules and Materials, Heyendaalseweg 135, NL-6525 AJ Nijmegen, The Netherlands}

\author{M. Sch\"uler}
\affiliation{Universit{\"a}t Bremen, Institut f{\"u}r Theoretische Physik, Otto-Hahn-Allee 1, 28359 Bremen, Germany}
\affiliation{Universit{\"a}t Bremen, Bremen Center for Computational Materials Science, Am Fallturm 1,
28359 Bremen, Germany}

\author{M. I. Katsnelson}
\affiliation{Radboud University, Institute for Molecules and Materials, Heyendaalseweg 135, NL-6525 AJ Nijmegen, The Netherlands}

\author{T. O. Wehling}
\affiliation{Universit{\"a}t Bremen, Institut f{\"u}r Theoretische Physik, Otto-Hahn-Allee 1, 28359 Bremen, Germany}
\affiliation{Universit{\"a}t Bremen, Bremen Center for Computational Materials Science, Am Fallturm 1,
28359 Bremen, Germany}

\begin{abstract}
We investigate the Peierls-Feynman-Bogoliubov variational principle to map Hubbard models with nonlocal interactions to effective models with only local interactions. We study the renormalization of the local interaction induced by nearest-neighbor interaction and assess the quality of the effective Hubbard models in reproducing observables of the corresponding extended Hubbard models. We compare the renormalization of the local interactions as obtained from numerically exact determinant Quantum Monte Carlo to approximate but more generally applicable calculations using dual boson, dynamical mean field theory, and the random phase approximation. These more approximate approaches are crucial for any application with real materials in mind. Furthermore, we use the dual boson method to calculate observables of the extended Hubbard models directly and benchmark these against determinant Quantum Monte Carlo simulations of the effective Hubbard model.
\end{abstract}

\maketitle

\section{Introduction}

Strongly correlated materials form one of the most challenging problems in condensed matter physics. Determining the electronic structure of a realistic material usually involves a multi-step process. First, the original problem is downfolded to a model system that describes the strongly correlated sector. Subsequently, an approximate solution for the model system is sought. 
The model system should be simple enough that it is computationally tractable. 
The Hubbard model of itinerant electrons with a local interaction is a popular choice~\cite{Hubbard63,Gutzwiller63,Kanamori63,Hubbard64,Gutzwiller64}.
The original electrons of the model, however, have a long-range Coulomb interaction, so in general one would prefer to use an extended Hubbard model that also includes the non-local interactions, if this was computationally feasible.

The Peierls-Feynman-Bogoliubov variational principle~\cite{Peierls38,Bogoliubov58,Feynman72} can be used to map extended Hubbard models with non-local interactions to effective models with only local interactions~\cite{schuler_optimal_2013}. Previously, this method has been used to estimate the effective local interaction in materials with a hexagonal lattice, such as graphene~\cite{schuler_optimal_2013}. Our focus here is on the square lattice Hubbard model with nearest-neighbor interaction. This system has been studied extensively as a testbed for theories developed over the last few years, using, e.g., Quantum Monte Carlo~\cite{zhang_extended_1989}, EDMFT+GW~\cite{Sun02,Ayral13,Huang14}, Dual Boson~\cite{van_loon_beyond_2014} and a slave boson approach \cite{Lhoutellier15,Fresard15}.
Of particular interest is how the renormalized effective interaction depends on the parameters of the Hubbard model. We show that the screening that leads to a renormalization of the local interaction depends strongly on interaction strength and filling.

We use four different computational methods, namely Determinant Quantum Monte Carlo~\cite{blankenbecler_monte_1981} (DQMC), Dual Boson~\cite{rubtsov_dual_2012,van_loon_beyond_2014,stepanov_self-consistent_2016} (DB), Dynamical Mean-Field Theory~\cite{metzner89,georges_dynamical_1996} (DMFT) and the Random Phase Approximation (RPA)~\cite{mahan00}. Two of these, DQMC and DMFT, are restricted to systems with local interaction. This is one of the main motivations for the effective Hubbard model approach: the variational principle allows predictions about the extended Hubbard model using computational methods that are restricted to the Hubbard model. 
Since DQMC is numerically exact, it provides the perfect benchmark for the other methods. 
DMFT, on the other hand, is an approximation that can be extended to realistic, multi-orbital systems. Coupled with DFT, it forms the workhorse of the strongly correlated materials community. 
With the variational principle, non-local interactions can be incorporated into DMFT calculations with relative ease.

RPA and DB allow us to do calculations in the extended Hubbard model. The RPA has only a limited range of validity, since it is a theory for weakly interacting systems. At the same time, computationally it is the simplest of all the theories considered here. DB, on the other hand, is an extension of DMFT that incorporates strong correlation effects. We use it to determine observables in the presence of non-local interaction effects, and to study which observables follow the predictions of the variational principle.

The remainder of this work is structured as follows:
In Sec.~\ref{sec:methods} we give a short overview of the variational principle used to determine the effective local interaction and of the methods used to obtain numerical results. In Sec.~\ref{sec:corrFuncs}, we determine the effective interaction strength in the half-filled Hubbard model. In Sec.~\ref{sec:benchmarkingobservables}, we use DQMC as benchmark for several observables calculated in DB, and in Sec.~\ref{sec:observables} we calculate these observables in the corresponding extended Hubbard model using DB. In Sec.~\ref{sec:alpha:doped} and \ref{sec:observables:doped}, we perform the same analysis for a strongly doped Hubbard model.

\section{Methods}
\label{sec:methods}

The extended Hubbard model on the square lattice with nearest-neighbor hopping reads
\begin{align}
H = -t\sum_{\langle i,j\rangle,\sigma} c_{i\sigma}^\dagger c^{\phantom{\dagger}}_{j\sigma} + U\sum_i n_{i\up}n_{i\dn} + \frac{1}{2} \sum_{\stackrel{i\neq j}{\sigma,\sigma'}}V_{ij} n_{i\sigma}n_{j\sigma'}, \label{eq:exHub}
\end{align}
where $t$ is the nearest-neighbor hopping-matrix element and $\langle i,j\rangle$ is a sum over nearest neighbors. $U$ and $V_{ij}$ are the local and non-local Coulomb matrix elements, respectively. The extended Hubbard model is a particular case of the so-called polar model that has been studied since the 1930's~\cite{Schubin34,Vonsovsky79_1,Vonsovsky79_2}.

We briefly review the main results of the variational method to map extended Hubbard models to \textit{effective} Hubbard models with only local interactions\cite{schuler_optimal_2013}. The effective Hubbard model, reading
\begin{align}
\tilde H = -t\sum_{\langle i,j\rangle,\sigma} c_{i\sigma}^\dagger c_{j\sigma} + \tilde U\sum_i n_{i\up}n_{i\dn},  \label{eq:effHub}
\end{align}
is varied with respect to the effective local interaction $\tilde U$ in order to minimize a free energy functional. This is equivalent to solving 
\begin{align}
\tilde U =U-\sum_{j\neq 0}V_{0j}  \frac{\partial_{\tilde U}\langle{n_{0}n_{j}}\rangle_{\tilde H}}{\partial_{\tilde U} \langle n_{0}n_{0}\rangle_{\tilde H}} \label{eq:ustar}
\end{align}
for $\tilde U$. For only nearest-neighbor interaction $V$, Eq. \ref{eq:ustar} simplifies to
\begin{align}
\tilde U =U-V  \alpha(\tilde U) \label{eq:ustarAlpha},
\end{align}
where we have introduced the nearest-neighbor renormalization strength 
\begin{align}
\alpha(\tilde U) = \sum_{\langle 0,j \rangle}  \frac{\partial_{\tilde U}\langle{n_{0}n_{j}}\rangle_{\tilde H}}{\partial_{\tilde U} \langle n_{0}n_{0}\rangle_{\tilde H}}. \label{eq:alpha}
\end{align}

Physically, this $\alpha$ describes the effective screening of the local interaction by the non-local interaction effects. $\alpha$ is a function of $\tilde U$, so it can be determined with knowledge of the effective local model alone. There is no assumption on the magnitude of any of the parameters in the original model. The only limit of applicability is that not all physical effects present in the extended Hubbard model can be captured in an effective local interaction, as discussed below.

The variational principle involves the calculation of charge correlation functions of the effective Hubbard model: $\langle n_i n_j \rangle_{\tilde H}$, where $n_i = n_{i\up} + n_{i\dn}$. 
In this work, we use several methods to calculate the charge correlation functions: The DQMC~\cite{blankenbecler_monte_1981}, RPA~\cite{mahan00}, DMFT~\cite{metzner89,georges_dynamical_1996} and DB ~\cite{rubtsov_dual_2012,van_loon_beyond_2014} methods. In the remainder of this section, we give a short summary of these methods and associated numerical details. This part can safely be skipped on the first read.
 
The DQMC method is numerically exact and has been used for the variational principle before~\cite{schuler_optimal_2013}. We use the implementation of the \textsc{quest} code\footnote{``QUantum Electron Simulation Toolbox'' \textsc{quest} 1.3.0 A. Tomas, C-C. Chang, Z-J. Bai, and R. Scalettar, (\url{http://quest.ucdavis.edu/})}. We obtain susceptibilities on Matsubara frequencies by Fourier transforming imaginary-time data. Disadvantages of DQMC are that it cannot be applied to the extended Hubbard model with non-local interaction straightforwardly \cite{zhang_extended_1989,PhysRevLett.68.353,golor_nonlocal_2015} and that it suffers from a sign problem away from half-filled systems. We perform the DQMC calculations for finite $12 \times 12$ (for half-filling) and $8\times 8$ (away from half-filling) systems with Trotter discretisation $\Delta \tau = 0.025$.

The RPA approach~\cite{mahan00} is geared towards weakly-interacting systems. 
In a noninteracting system, the susceptibility is given by a ``bubble'' diagram of two Green's functions,
\begin{align}
\chi_0(i\omega_n,\mathbf{q}) =& -\frac{1}{\beta} \sum_{\nu_m} [G^0 G^0]_{\nu\omega \mathbf{q}},\notag \\
[G^0 G^0]_{\nu\omega \mathbf{q}} =& \frac{1}{N} \sum_{\mathbf{k}} G^0(i\nu_m+i\omega_n,\mathbf{k}+\mathbf{q}) G^0(\nu_m,\mathbf{k}),
\end{align}
where $G^0(\nu_m,\mathbf{k})$ is the noninteracting Green's function, $N$ is the number of $\mathbf{k}$ points, and $\nu_m,\omega_n$ are fermionic and bosonic Matsubara frequencies, respectively.
Then, the susceptibility of the (weakly) interacting system is
\begin{align}
\chi^{-1}_\text{RPA}(i\omega_n,\mathbf{q}) = \chi^{-1}_0(i \omega_n,\mathbf{q}) + U+V(\mathbf{q}), \label{eq:RPA}
\end{align}
where $V(\mathbf{q})$ is the Fourier transform of $V_{ij}$ in Eq.~\eqref{eq:exHub}.
Finally, the (equal-time) charge correlation functions are obtained by a Fourier transform to real space and by summing over the Matsubara frequency $\omega_n$:
\begin{align}
\langle n_{0}n_{i}\rangle - \langle n_0\rangle \langle n_i \rangle = \frac{2}{\beta N}\sum_n \sum_\mathbf{q} e^{i\mathbf{q} \mathbf{r}_i}\chi_\text{RPA}(i\omega_n,\mathbf{q}).
\end{align}

Dynamical mean-field theory~\cite{georges_dynamical_1996} is an approximate method that includes \emph{local} correlation effects. The method applies to systems with local interactions only, i.e., $V=0$. The approach is based on a self-consistently determined auxiliary single-site problem. 
As in RPA, we obtain the correlation functions from the susceptibility in momentum and frequency space, which is given by
\begin{align}
 \chi_{\text{DMFT}}^{-1}(i\omega_n,\mathbf{q}) = -\hat{[\mathcal{G}\mathcal{G}]}(i\omega_n,\mathbf{q})^{-1} - \hat{\Gamma}_{\omega_n}.
\end{align}
Here $\mathcal{G}$ is the DMFT Green's function, which has a local self-energy, $\Gamma$ is the particle-hole irreducible two-particle vertex of the auxiliary single-site problem. $[\mathcal{G}\mathcal{G}]$ is shorthand for the product of two Green's functions. The equation has a matrix structure in fermionic frequencies~\cite{georges_dynamical_1996,PavariniJulich}, as indicated by the hats, which we have suppressed for notational convenience.

The DB method~\cite{rubtsov_dual_2012} is a diagrammatic extension of DMFT that allows for the treatment of non-local interactions directly, via an effective frequency dependent interaction $U(i\omega_n)$. We apply self-consistent DB~\cite{stepanov_self-consistent_2016} in the charge and magnetic ($S^z$) channel to obtain consistent correlation functions~\cite{van_loon_double_2016}. In DB, the expression for the susceptibility is 
\begin{align}
\chi_{\text{DB}}^{-1}(i\omega_n,\mathbf{q}) = \chi_{\text{DMFT}}^{-1}(i\omega_n,\mathbf{q}) + U+V(\mathbf{q})- U(i\omega_n), \label{eq:susc:db}
\end{align}
with the important caveat that $\chi_{\text{DMFT}}$ is determined using the DB auxiliary single-site problem, so that the $\mathcal{G}$ and $\Gamma$ that enter this equation are different from the ones in DMFT.
DMFT is recovered when $V(\mathbf{q})=0$ and $U=U(i\omega_n)$.

The DB calculations are performed on a $64\times 64$ lattice. The DB implementation and the CT-HYB~\cite{Werner06} impurity solver are based on the ALPS libraries~\cite{ALPS2,Hafermann13}. The impurity solver takes into account retarded interactions~\cite{Ayral13} and uses improved estimators for the two-particle quantities~\cite{Hafermann14}.
We use converged extended Dynamical-Mean Field Theory (EDMFT, c.f. Appendix~\ref{app:edmft}) calculations as the starting point for the DB self-consistency. 
In Appendix~\ref{app:uw}, we show the converged dynamic interaction $U(i\omega_n)$ in both EDMFT and DB.
The number of iterations needed to achieve the self-consistent hybridization and dynamic interaction increases from less than ten to approximately forty between $U/t=7$ and $U/t=8$. In this region, the vertex corrections to the susceptibility are very strong~\cite{van_loon_double_2016}, and $U(i\omega_n)$ converges slowly. This issue makes calculations at higher $U$ very expensive computationally, and for that reason most of the DB calculations are at $U/t<8$.

\section{Effective local interaction at half-filling} 
\label{sec:corrFuncs}

Results from all methods presented in this work are obtained at the temperature $\beta t=2$. We discuss the temperature dependence of our results at the end of this section. We use $t=1$ as the unit of energy.

\begin{figure}[b]
\begin{center}
\includegraphics{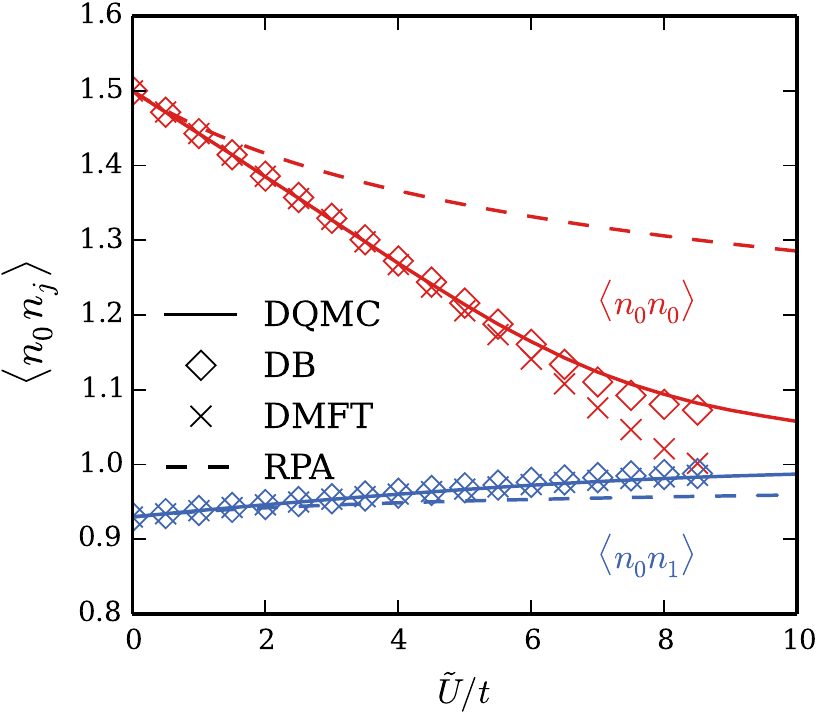}
\end{center}
\caption{(Color online) Local (red) and nearest-neighbor (blue) charge correlation functions of half filled nearest-neighbor hopping Hubbard model on a square lattice obtained from DQMC (full line), DB (diamonds), DMFT (crosses), and RPA (dashed line).}
\label{fig:corrFuncHalf}
\end{figure}

To begin with, we discuss the half filled, $\langle n_0\rangle=1.0$, extended Hubbard model with only nearest-neighbor interaction terms. Therefore, the evaluation of Eq. \ref{eq:ustar} involves the calculation of the local and nearest-neighbor charge correlation functions. The results from DQMC, DB, DMFT and RPA are depicted in Fig. \ref{fig:corrFuncHalf} for interaction strengths up to $U/t=10$. As expected, all methods reproduce the non-interacting case exactly. Wick's theorem theorem applies to the non-interacting system, so $\langle n_0 n_0 \rangle=\langle n_0 \rangle + 2\langle n_{0\uparrow} \rangle \langle n_{0\downarrow} \rangle $ and $\langle n_0 n_1 \rangle=\langle n_0 \rangle\langle n_1 \rangle - \sum_{\sigma}\langle c^\dagger_{0\sigma}c_{1\sigma} \rangle \langle c^\dagger_{1\sigma}c_{0\sigma} \rangle $. The DQMC results approach the strong coupling result for $U\gg t$, where every site has one exactly electron and $\langle n_0 n_0 \rangle=\langle n_0 \rangle=1$ and $\langle n_0 n_1 \rangle  =\langle n_0 \rangle \langle n_1 \rangle=1$.   For intermediate interaction strengths ($U/t \lesssim 5$), results from DQMC, DMFT and DB are virtually indistinguishable on this scale. For larger interactions, differences between the exact DQMC and both the DMFT and the DB approximations are visible. Clearly, the DB method improves the DMFT results. The RPA results are considerably off the DQMC results, especially at larger interaction strengths. We note that the agreement between the methods is considerably better for the nearest-neighbor correlation function than for the local correlation function.
 
\begin{figure}[htb]
\begin{center}
\includegraphics{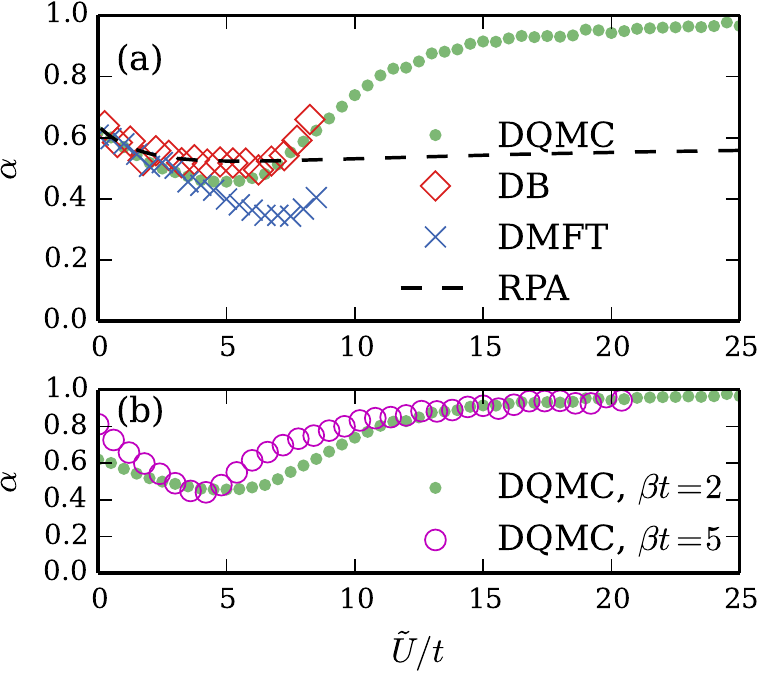}
\end{center}
\caption{(Color online) Nearest-neighbor renormalization strength $\alpha(\tilde U)$ of the half-filled nearest-neighbor hopping Hubbard model on a square lattice. (a) $\alpha(\tilde U)$ at $\beta t = 2$, as obtained from DQMC (green dots), DB (red diamonds), DMFT (blue crosses) and RPA (dashed line). (b) At $\beta t=2$ (green dots) and $\beta t = 5$ (magenta circles) using DQMC.}
\label{fig:alphaHalf}
\end{figure}

Next, we consider the nearest-neighbor renormalization strength $\alpha(\tilde U)$, depicted in Fig. \ref{fig:alphaHalf} (a), calculated from Eq. \ref{eq:alpha} with the different approximations. The simplest approximation to Eq. \ref{eq:ustar} discussed in Ref. \onlinecite{schuler_optimal_2013} leads to $\tilde U = U - V$, i.e., a constant $\alpha(\tilde U) = 1$. 
This approximation can be derived by assuming that the correlation between sites that are not nearest neighbors is zero, i.e., that the system is very strongly localized.

The DQMC result indeed shows that the approximation $\alpha(\tilde U) = 1$ is only valid for $\tilde U\gg t$. In this limit, sites that are more than one lattice spacing apart are uncorrelated, which is sufficient to prove $\alpha=1$~\cite{schuler_optimal_2013}.
In fact, $\alpha(\tilde U)$ has a minimum at intermediate $\tilde U$ before increasing towards 1. DB agrees quite well with the exact DQMC results, with the largest deviations occurring around the minimum of $\alpha$. For $\tilde U/t>5$, DMFT starts to deviate from DQMC. The DMFT results do show a minimum, however, that minimum is located at slightly larger $\tilde U$. For small interactions ($\tilde U/t \lesssim 5$) the RPA follows the behavior found in DQMC. However, RPA does not reproduce  the minimum found around $U/t\sim 5$ and  is off by a factor of almost 2 at large interactions.

To study the role of temperature, we have also done DQMC calculations at $\beta t = 5$ instead of $\beta t = 2$. There, $\alpha(\tilde{U})$ is qualitatively similar, as visible in Fig. \ref{fig:alphaHalf}(b). At $\tilde{U}=0$, the nearest-neighbor renormalization strength $\alpha$ is larger, as $\tilde{U}$ increases it goes to a slightly deeper minimum that occurs at smaller $\tilde{U}$, and finally for large $\tilde{U}$ the renormalization strength goes towards 1.

To give a sense of scale, we remind the reader that $\beta t =5$ and $\tilde{U}/t=4$ is in a region where antiferromagnetic fluctuations are strong. This point is close to the metal-insulator transition according to a combination of diagrammatic and Monte Carlo techniques~\cite{Schafer15} and also to the DMFT N\'eel temperature. The (single-site) DMFT Mott metal-insulator transition, on the other hand, occurs at higher interaction strength $\tilde{U}/t=10$ -- $12$. 

\section{Benchmarking DB observables}
\label{sec:benchmarkingobservables}

The variational principle only deals with the free energy. 
In practical calculations, the main interest often lies with other observables, such as the Green's function and the double occupancy. The question is how well the optimal $\tilde U$ Hubbard model reproduces the observables of the original extended Hubbard model with parameters $U$ and $V$. We want to use DB to calculate the observables of the extended Hubbard model. Before we do that, we study the accuracy of DB at $V=0$, where we can use DQMC as a benchmark. 

\begin{figure}
\includegraphics{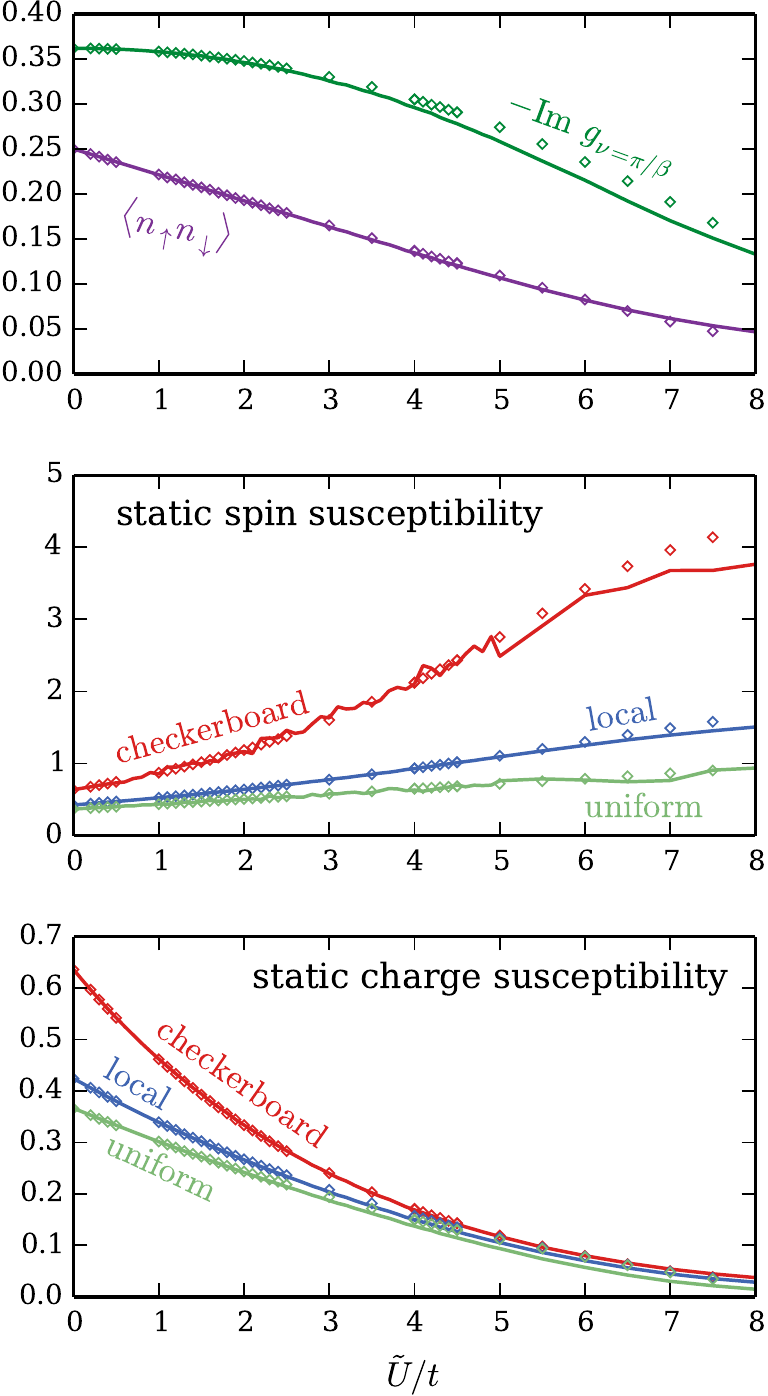}
 \caption{
  Observables in the Hubbard model ($V=0$) obtained using DQMC (lines) and the DB method (diamonds). 
 }
 \label{fig:observables2}
\end{figure}

In the previous section, we have seen that the DB results for the local and nearest-neighbor correlation function are accurate. In Fig.~\ref{fig:observables2}, we extend this conclusion to other observables, by comparing them to the DQMC values. 
In Fig.~\ref{fig:observables2}(a), we show the imaginary part of the local Green's function at the lowest Matsubara frequency and the double occupancy. The latter is equal to $\av{n_0 n_0}/2 - 1/2$, cf. Fig.~\ref{fig:corrFuncHalf}. 
In Fig.~\ref{fig:observables2}(b), we show some zero (Matsubara) frequency spin susceptibilities. These provide insight into the response of the system to static external fields.
They are natural observables for DB, since we calculate the entire momentum and frequency dependent susceptibility according to Eq.~\eqref{eq:susc:db}. 
In DQMC, the susceptibility is determined as a function of imaginary time and we obtain the static component by a Fourier transform. 
In the figure, we show the $q=(\pi,\pi)$ (checkerboard), $q=(0,0)$ (uniform) and $q$-averaged (local) spin susceptibility.
In Fig.~\ref{fig:observables2}(c), we do the same for the charge susceptibility.
For all the observables shown, we find a good qualitative and quantitative agreement between DQMC and DB. Deviations start to set in at interaction strengths $\tilde{U}/t\gtrsim 6$. 

Fig.~\ref{fig:observables2} shows that the local repulsion $\tilde{U}$ suppresses the double occupancy and charge excitations in general, whereas spin excitations are enhanced. The checkerboard spin susceptibility, corresponding to antiferromagnetism, increases the most. We should note that the temperature studied here, $\beta t = 2$, is above the Mott transition temperature, and all observables depend smoothly on $\tilde{U}$.

\section{Observables at finite $V$}
\label{sec:observables}

\begin{figure*}
\includegraphics[]{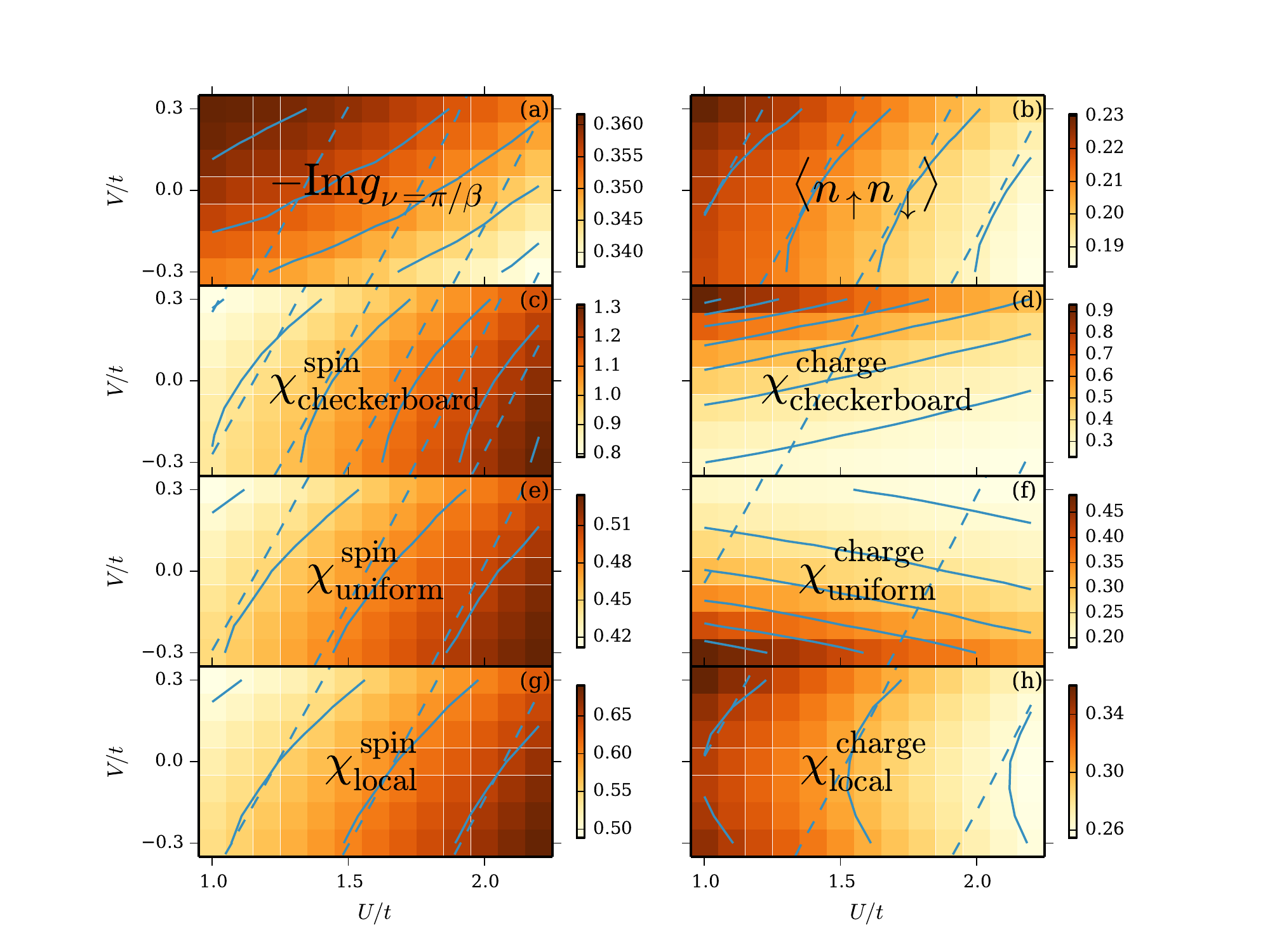}
 \caption{
  Observables in the extended Hubbard model obtained using the DB method. Every colored square represents a DB calculation, they are separated by $\Delta U=0.1$ and $\Delta V=0.1$. The solid lines show isolines where the observable is constant, the dashed lines show lines of constant $\tilde U$ according to DQMC.
  The constant values of the isolines correspond to the tick labels in the colorbars.
 }
 \label{fig:observables}
\end{figure*}

Now that we have confidence in the predictions of DB, we can use them as a benchmark for the variational principle at finite values of $V$.
To study this, we have done DB calculations for $1.0 \leq U/t \leq 2.2$ and $-0.3 \leq V/t \leq 0.3$, the results of which are shown in Fig.~\ref{fig:observables}. The colored plots give the value of the observable according to DB, with each colored square corresponding to a specific value of $U$ and $V$. The solid lines are approximate isolines in the $(U,V)$-plane along which the observable is constant. The dashed lines, on the other hand, indicate constant $\tilde U$. If the variational principle were exact, these would be the isolines of the observables.

Fig.~\ref{fig:observables}(a) contains the value of the Green's function on the first Matsubara frequency $\pi/\beta$, $-\Im g_{\nu=\pi/\beta}$. The DB observables does show a roughly linear dependence on $V$, albeit with a different slope than the variational principle predicts.

Next, in Fig.~\ref{fig:observables}(b), for the double occupancy, there is a very good match between the DB calculations and the variational scheme. This correspondence is not accidental, the variational principle gives the exact double occupancy to first order in $V$, as is shown in Appendix \ref{app:smallV}. The double occupancy is a special operator in this context, since it is directly connected to the variational parameter $\tilde{U}$.
This explains the matching tangents at $V=0$ in Fig. 4(b). For larger $\left|V\right|$, the curvature in the DB results shows a divergence from the simple $\tilde U = U-\alpha V$ prescription. As $V$ becomes large compared to the Hubbard parameters $\tilde U$ and $t$, it is no longer reasonable to expect the effective Hubbard model to do a good job in describing the relevant physics.

In Fig.~\ref{fig:observables}(c), (e) and (g), we move to the zero-frequency spin susceptibilities, namely the checkerboard [$q=(\pi,\pi)$], uniform [$q=(0,0)$] and  local part [$q$-average] of the spin susceptibility, respectively. All three show a reasonable, though not perfect, match between the variational principle prediction and the DB results.

The corresponding correlation functions in the charge sector, in Fig.~\ref{fig:observables}(d), (f) and (h), show a very different dependence on $V$. The checkerboard correlation function does have a linear dependence on $V$, with very small slope. In the uniform charge susceptibility, the sign of the $V$-dependence has changed, and for the local susceptibility the dependence is even quadratic instead of linear. 
This poor match is not a surprise. The non-local interaction $V$ directly and explicitly enters the charge dynamics, as in Eq.~\eqref{eq:susc:db}, and the effective local interaction can only give a poor description of that dependence.

As an alternative to DB, it is also possible to use EDMFT to do calculations at finite $V$. In Appendix~\ref{app:edmft}, we show the resulting observables. EDMFT has the advantage of being simpler, however the approximate treatment of the momentum structure of correlation functions leads to quadratic instead of linear scaling in $V$, and a poor description of the spin susceptibility, as shown in the Appendix.

\begin{figure}[htb]
\begin{center}
\includegraphics{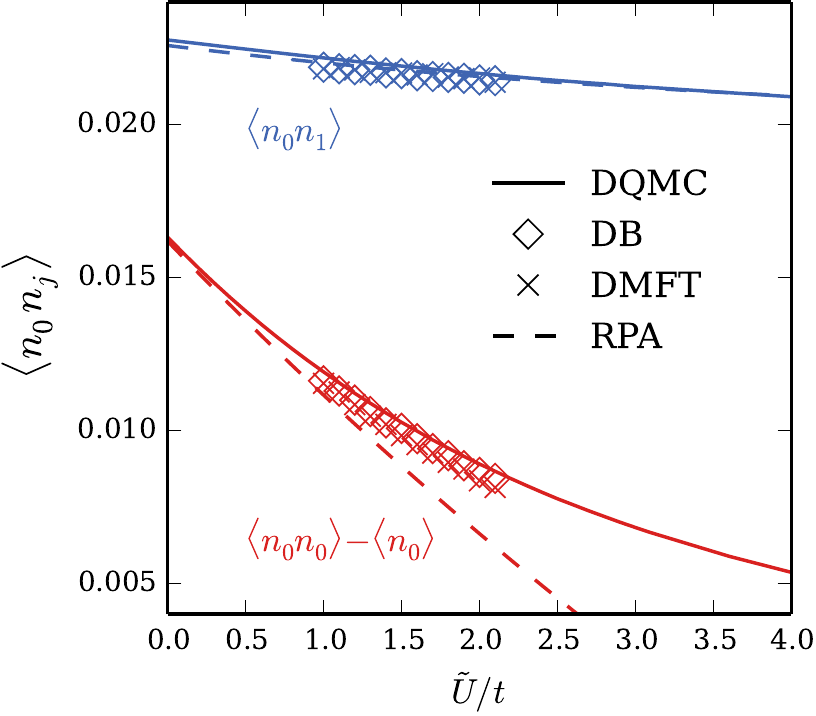}
\end{center}
\caption{(Color online) Local (red) and nearest-neighbor (blue) charge correlation functions of the hole doped ($\langle n_0\rangle=0.18$) nearest-neighbor hopping Hubbard model on a square lattice obtained from DQMC (full line), DB (diamonds), DMFT (crosses), and RPA (dashed line). }
\label{fig:corrFuncDope}
\end{figure}

\section{Effective local interaction away from half-filling}
\label{sec:alpha:doped}

In order to study the performance of the different approximations as well as the renormalizations in dependence of the filling, we study the same Hubbard model as above with a filling of $\langle n_0\rangle=0.18$, well below the optimal filling for high-$T_c$ superconductors, of $\langle n_0 \rangle \approx 0.8$. At this small filling, the DQMC sign problem is not very severe and computations are feasible. In addition, we restrict ourselves to intermediate interaction strengths.

We follow the same approach as before and start by determining the charge correlation functions.
These are shown in Fig. \ref{fig:corrFuncDope}. We come to very similar conclusions as in the case of half filling. DQMC, DB, and DMFT results agree closely in the investigated regime. RPA agrees well for the nearest-neighbor correlation function, and poorly for the local correlator.

\begin{figure}[htb]
\begin{center}
\includegraphics{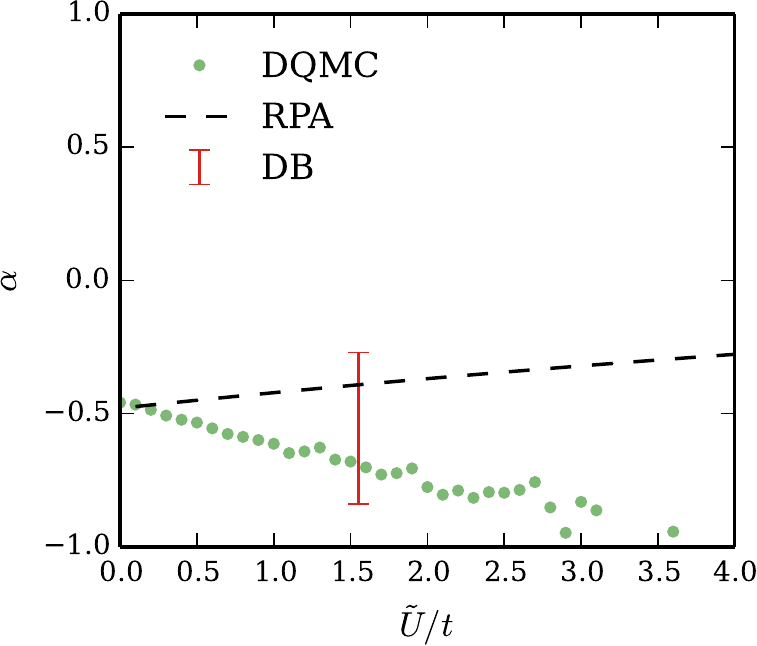}
\end{center}
\caption{(Color online) Nearest-neighbor interaction strength $\alpha(\tilde U)$ of the hole doped ($\langle n_0\rangle=0.18$) nearest-neighbor hopping Hubbard model on a square lattice obtained from DQMC (green dots), DB (red error bar), and RPA (dashed line). }
\label{fig:alphaDope}
\end{figure}

DQMC, DMFT and DB all operate in the grand-canonical ensemble, at fixed chemical potential $\mu$. The results at fixed density are obtained by interpolating between simulations at fixed chemical potential. This interpolation step introduces additional uncertainty into the determination of the correlators. This is especially visible when taking the numerical derivative of the correlation functions to obtain $\alpha$, since the difference quotient is very susceptible to noise. To estimate $\alpha$ for DB, we used a linear fit through all the data points in Fig.~\ref{fig:corrFuncDope}, and then we used the linear coefficients of these fits to determine $\alpha$. The resulting $\alpha$ is shown Fig.~\ref{fig:alphaDope}. To obtain an error estimate, a quadratic fit of the data in Fig.~\ref{fig:corrFuncDope} was done, this results in $\tilde{U}$-dependent derivatives and the spread in the derivates was used to determine the error bar. DMFT (not shown) gives a result within the (rather large) error bars of DB.

The first thing that is clear is that $\alpha \neq 1$, so the simple formula $\tilde{U}=U-V$ does not hold.
Interestingly, for this hole doped case, the nearest-neighbor renormalization strength is even negative, i.e., the nonlocal interaction increases the effective local interaction, $\alpha<0$. This is in line with findings in the context of doped benzene models in Ref. \onlinecite{schuler_optimal_2013} and can be understood in terms of Wigner crystallization~\cite{wigner34}. 
In a very empty system, the local interaction $\tilde{U}$ suppresses not only the probability to find a second electron on the same site, but also in the vicinity of the first electron.
The effect of the non-local interaction $V$ is also to keep electrons away from each other, so a positive $V$ leads to a larger effective $\tilde{U}$.
The DQMC results show that this renormalization increases for larger interaction. While RPA gives the correct sign, it underestimates $|\alpha|$ at all finite interaction strengths and it predicts a decreasing renormalization for growing interaction, which is the wrong trend.

\section{Observables in doped systems}

\label{sec:observables:doped}

\begin{figure}
\includegraphics{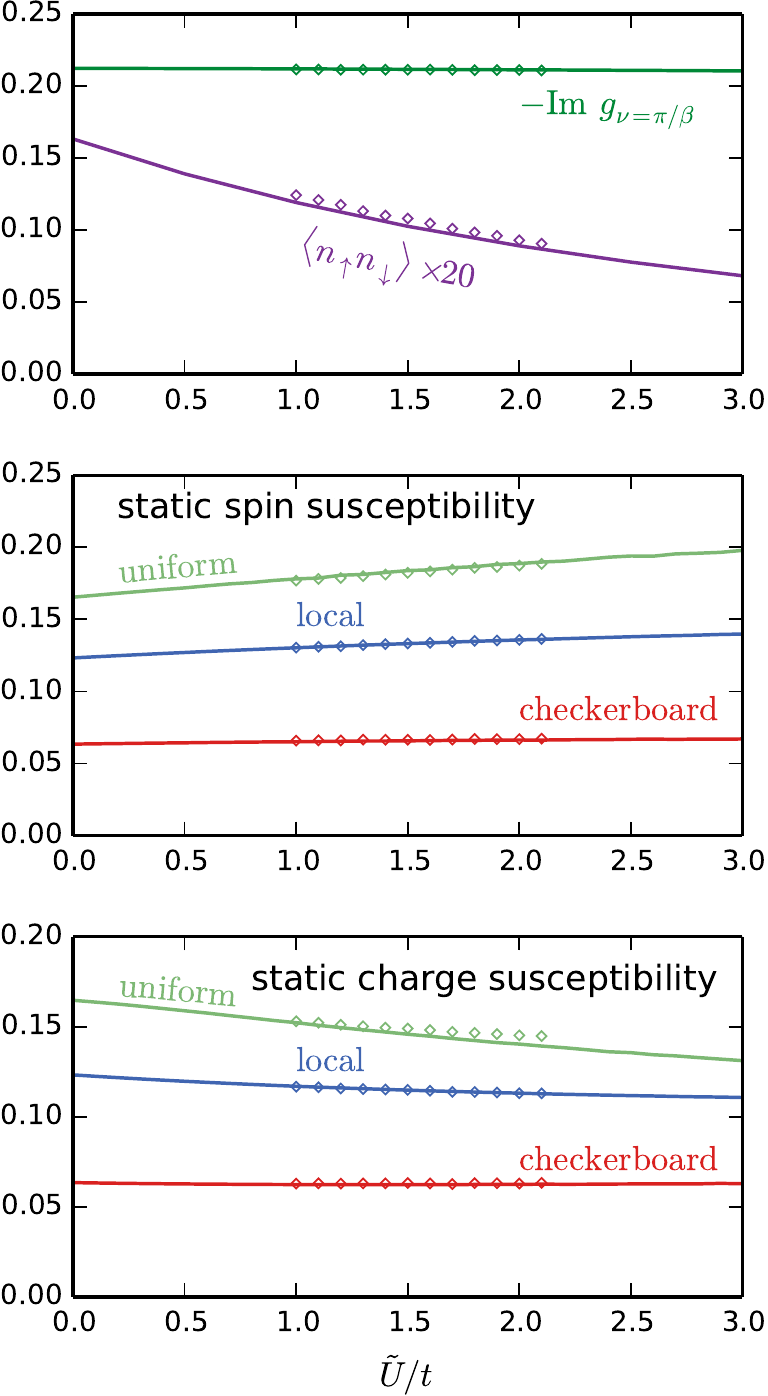}
 \caption{
  Observables in the Hubbard model away from half-filling ($\av{n_0}$ and $V=0$) obtained using DQMC (lines) and the DB method (diamonds), cf. Fig.~\ref{fig:observables2} for the half-filled system. 
 }
 \label{fig:doped:observables2}
\end{figure}

\begin{figure*}
\includegraphics[]{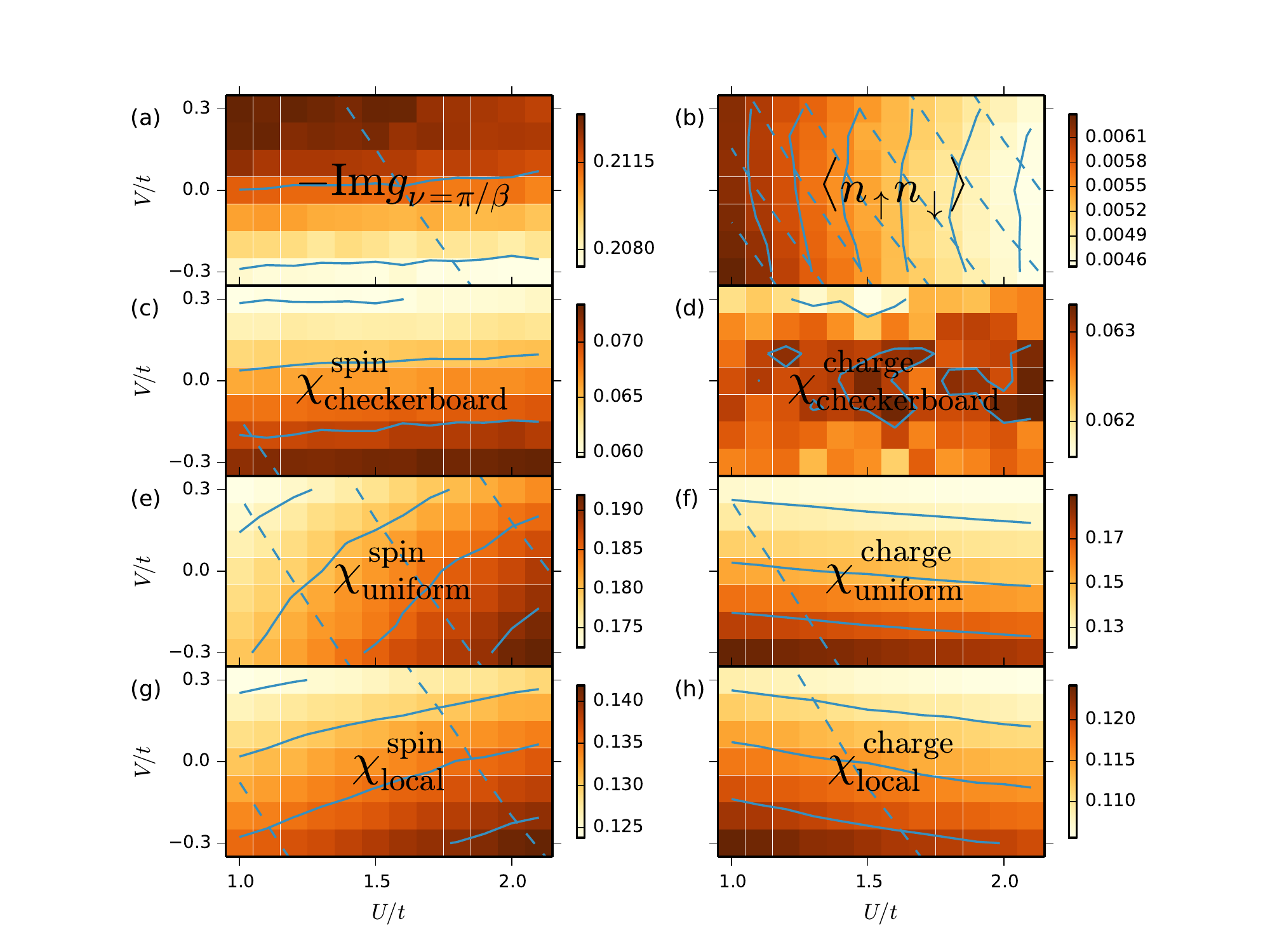}
 \caption{
  Observables in the extended Hubbard model, away from half-filling at $\av{n_0}=0.18$, obtained using the DB method. See Fig.~\ref{fig:observables_doped} for the half-filled system.
 }
 \label{fig:observables_doped}
\end{figure*}

Finally, we also study the observables of the doped system. 
As before, we start by comparing DQMC and DB observables at $V=0$, this is shown in Fig.~\ref{fig:doped:observables2}. We again find a good match between the DQMC and DB results. Secondly, we show the observables at finite $V$ in Fig.~\ref{fig:observables_doped}. As before, the isolines predicted by DQMC are shown as dashed lines. Here, however, they do not match at all with the observables from DB.

Physically, the strongly doped $\av{n_0}=0.18$ system is very different from the half-filled Hubbard model. 
The local interaction $U$ only affects electron pairs that occupy the same site, and as a result, an observable like $g_{\nu=\pi/\beta}$ depends only very weakly on $U$, as seen in Fig.~\ref{fig:doped:observables2}. This weak dependence holds even at higher values of $\tilde{U}/t$ (not shown).
As mentioned above, the local and non-local interaction creates precursors to Wigner crystallization~\cite{wigner34}.
Even in the non-interacting system, the probability to find two electrons at the same site is $\av{n_0}^2/4 < 1\%$. 
As such, it is very difficult to encapsulate the effect of $V$ into an effective $\tilde{U}$. Since the physics of the strongly-doped extended Hubbard model is not Hubbard-like, the effective mapping is not able to provide relevant results.

The susceptibility provides another clear difference between Figs.~\ref{fig:observables} and \ref{fig:observables_doped}. In the half-filled model, perfect nesting of the Fermi surface with the checkerboard wavevector $q=(\pi,\pi)$ creates a tendency towards checkerboard ordering (antiferromagnetic ordering in the spin channel) at lower temperatures. This is visible in the large values of the checkerboard susceptibilities in Fig.~\ref{fig:observables}. At the much lower density $\av{n_0}=0.18$, checkerboard ordering is not favored, and the checkerboard susceptibility is smaller than the uniform and local susceptibilities, and it depends only weakly on the interaction strength.

DMFT and DB both use a single-site auxiliary problem as the starting point of their approach. In this way, they are able to incorporate strong local correlation effects. Local correlation requires two particles at the same site, so this is expected to be less important at strong doping.
Based on this, we can expect the non-local correlations that DB includes on top of DMFT to become more important at strong doping~
\cite{van_loon_thermodynamic_2015,van_loon_double_2016}, and we can also expect that the local correlation effects included in DB do not significantly improve on simpler theories like the $GW$-method.

This almost empty system clearly requires an explicit treatment of the nonlocal interactions. The extended Hubbard model is physically very different from the purely local Hubbard model, so any attempt to use the variational principle to relate the two is ill-fated.

\section{Conclusions and Discussion}

We have studied the mapping of the extended Hubbard model onto effective local Hubbard models using a variational principle. In the half-filled Hubbard model, the simple prescription $\tilde{U} = U - V$ is only applicable at very high values of $U$. At intermediate $U$, the effective renormalization of the local interaction $\alpha$, with $\tilde{U} = U - \alpha V$, is reduced by as much as a factor of 2.

To determine the effective interaction, the local and nearest-neighbor correlation function of the Hubbard model are needed.
We find that the self-consistent DB approximation accurately reproduces the numerically exact DQMC results for the correlators, so that even the numerical derivatives come out similarly. 
DMFT performs qualitatively correct but quantitatively slightly worse at larger interaction strengths $\tilde{U}/t>5$, and misses the location of the minimum of $\alpha$.
RPA obtains the correct nearest-neighbor renormalization strength at small interaction, however it does not have a minimum and fails in the limit of large interaction.

For the nearly empty system (i.e. heavy hole doping, $\av{n_0}=0.18$), we find that non-local repulsion actually predicts a larger effective local interaction. Numerical calculations are more difficult in this parameter regime, which makes it difficult to assess exactly how well DMFT and DB match with the DQMC results. 

We have also studied how observables of the extended Hubbard model behave as a function of $V$, again using the DB approach. The effective Hubbard model predictions work well for most observables at half-filling. For charge correlation functions, though, the effective Hubbard model does not match with the finite $V$ results. This was to be expected, since the charge correlations depend explicitly on $V$. 

Away from half-filling, the match between observables in DB and in the effective Hubbard model is much worse. This parameter regime is dominated by Wigner crystallization physics, which is difficult to capture using the variational principle.

The contrasting behavior in these two scenarios teaches us that the mapping to an optimal local Hubbard model only has a chance to succeed when the physics of the system is essentially Hubbard-like. In the very empty system, where doubly occupied sites are rare, only changing the effective Hubbard parameter is insufficient to recover the Wigner localization physics. Similarly, the charge susceptibility in the half-filled system, which is directly driven by the non-local interaction, is not captured in the effective model.
On the other hand, the extended Hubbard model at half-filling is sufficiently similar to the local Hubbard model that a renormalization of the interaction strength suffices to explain the Green's function, double occupancy and spin susceptibility.

\acknowledgments

E.G.C.P. v. L. and M.I.K. acknowledge support from ERC Advanced Grant 338957 FEMTO/NANO. M.S. and T.O.W. acknowledge support from the Deutsche Forschungsgemeinschaft through the research unit 1346 and the University of Bremen through the Zentrale Forschungsf{\"o}rderung.
 
\appendix

 \section{Exact small-$V$ coefficient of the double occupancy}
 
 \label{app:smallV}
 
 The Peierls-Feynman-Bogoliubov variational principle suggests to use an effective local interaction $\tilde{U}$ to describe the effect of a nonlocal interaction $V$, with the prescription
 \begin{align}
  \tilde{U} =& U - \alpha V, \notag \\
  \alpha =& - \sum_{j \text{ nn of } 0}  \frac{\partial_{\tilde{U}} \av{n_{0} n_{j}}}{\partial_{\tilde{U}} \av{n_{0\up} n_{0\dn}} },
  \label{eq:app:alpha}
 \end{align}
 for that effective interaction, where we have restricted ourselves to nearest-neighbor interaction $V$. 
 
 For infinitesimally small nonlocal interaction $V=dV$, the change in effective local interaction $U-\tilde{U} = \alpha dV = dU$ will also be small, and
 \begin{align}
  \frac{dU}{dV} = \alpha.
 \end{align} 
 
 The expression for $\alpha$ contains two correlation functions, and both can be written as a derivative of the free energy,
 \begin{align}
  \sum_{j \text{ nn of } 0}\av{n_{0\up} n_{j\sigma'}} =& - \partial_{V} F  \\
  \av{n_{0\up} n_{0\dn}}     =& - \partial_U F. \label{eq:app:dbl}
 \end{align}
 This allows us to write Equation \eqref{eq:app:alpha} in terms of second derivatives of the free energy. Assuming sufficient continuity of the free energy, the order of the partial derivatives can be changed, 
\begin{align}
  \alpha =& - \frac{\partial_{V} \partial_{U} F}{\partial_U \partial_{U} F }, \label{eq:app:alpha2}
\end{align}
 where we have replaced $\partial_{\tilde{U}}$ by $\partial_{U}$ since $\tilde{U} - U$ is infinitesimal.
 
Now, we are interested in the value of the double occupancy, Equation \eqref{eq:app:dbl}, as a function of $U$ and $V$. Writing $\av{n_{0\up} n_{0\dn}}=A(U,V)=-\partial_U F$, we can expand around $V=0$.
\begin{align}
 A(U,dV) =& A(U,0) + dV \, \partial_V A \notag \\
 A(U-dU,0) =& A(U,0) + dU \, \partial_U A, 
\end{align}
so the optimal $\tilde{U}=U-dU$ is given by
\begin{align}
 A(U,dV) =& A(U-dU,0) \notag \\
 dV \, \partial_V A =& d U \, \partial_U A \notag \\
 \frac{dU}{dV} =& - \frac{\partial_V A}{\partial_U A} \notag \\
 =& - \frac{\partial_V \partial_U F}{\partial_U \partial_U F}
\end{align}
This is exactly Equation \eqref{eq:app:alpha2}. 

From this, we conclude that the Peierls-Feynman-Bogoliubov variational principle exactly reproduces the linear dependence on $V$ of the double occupancy. The only required assumption is that the free energy is sufficiently smooth to allow the interchange of partial derivatives.

We stress that this proof only works for the double occupancy, not for other observables. The reason for this is that the double occupancy is exactly the observable obtained by deriving the free energy with respect to the variational parameter $U$.
 
\section{EDMFT}
\label{app:edmft}

\begin{figure*}
\includegraphics[]{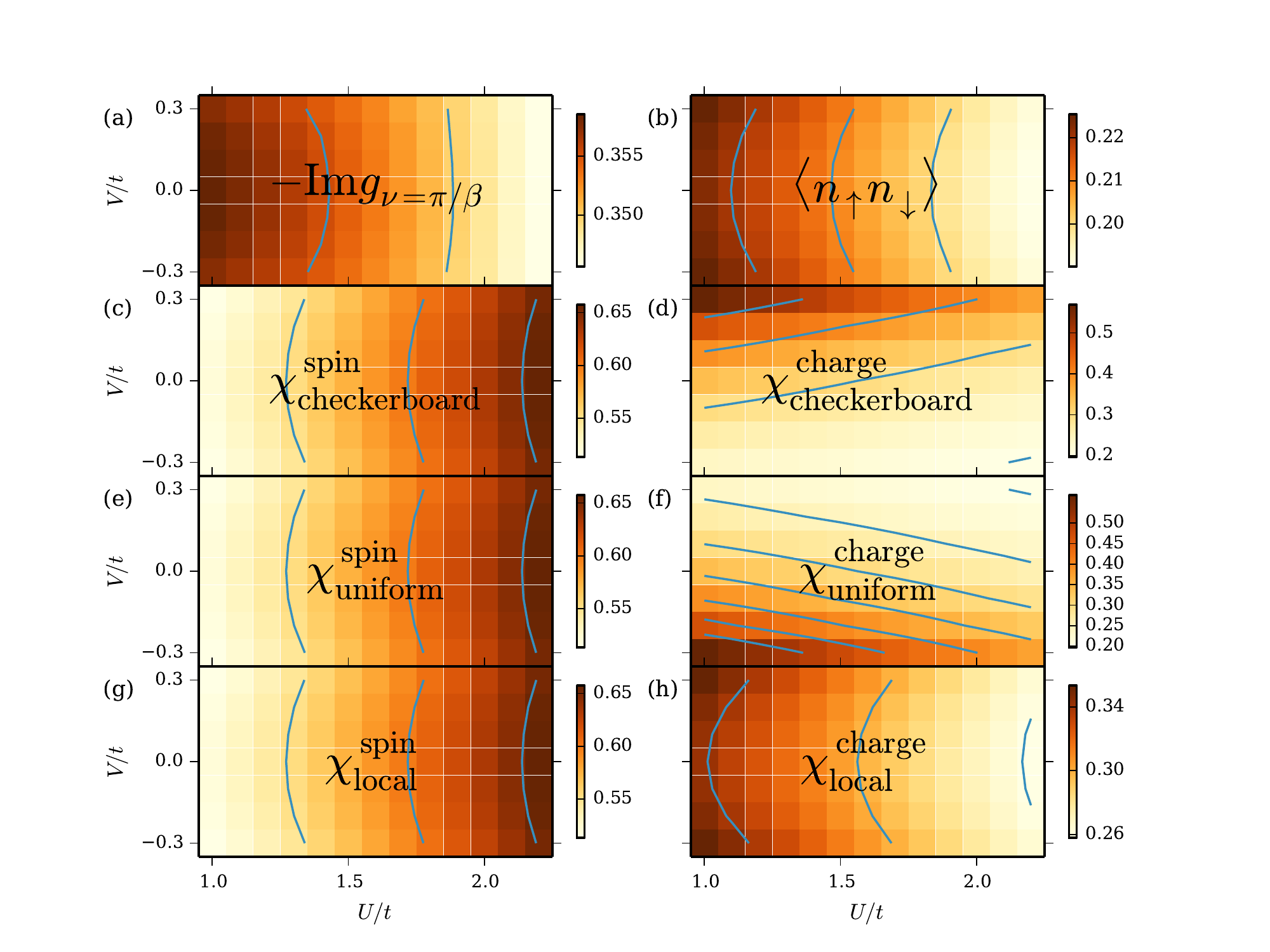}
 \caption{
  Observables in the extended Hubbard model obtained using the EDMFT method. Compare to the DB results in Fig.~\ref{fig:observables}. The solid lines show isolines where the observable is constant.
  The constant values of the isolines correspond to the tick labels in the colorbars.
 }
 \label{fig:observables:edmft}
\end{figure*}

DB is not the only DMFT-based approach that can incorporate nonlocal interactions. The simplest is EDMFT~\cite{Si96,Kajueter96,Smith00,Chitra00,Chitra01}, in which the susceptibility is determined as
\begin{align}
\chi_{\text{EDMFT}}^{-1}(i\omega_n,\mathbf{q}) = \chi_{\text{impurity}}^{-1}(i\omega_n) + U+V(\mathbf{q})- U(i\omega_n), \label{eq:susc:edmft}
\end{align}
where  $U(i\omega_n)$ is determined self-consistently similar to the DB approach. The difference between Eq.~\eqref{eq:susc:edmft} and Eq.~\eqref{eq:susc:db} is in the first term. EDMFT uses the momentum-independent impurity susceptibility as the starting point for calculating correlation functions. In particular, this means that the EDMFT susceptibility is independent of momentum whenever $V(\mathbf{q})$ is zero. In the context of Fig.~\ref{fig:observables2}, EDMFT predicts the susceptibility in a single channel to be the same at the various momenta (red, blue, and green lines in Fig.~\ref{fig:observables2}).
The figure shows that this is a good approximation only in the charge sector and at large $\tilde{U}$. Indeed, in this regime, DB and EDMFT give similar results~\cite{van_loon_beyond_2014}.

In Fig.~\ref{fig:observables:edmft} we show the EDMFT observables at finite $V$, similar to the DB results in Fig.~\ref{fig:observables}. The first thing to note are the different scales. This is particularly clear in the spin susceptibilities. The checkerboard, uniform and local spin susceptibility are identical in EDMFT according to Eq.~\eqref{eq:susc:edmft}, since $V(\mathbf{q})=0$ in the spin channel. This prediction of EDMFT is clearly inconsistent with the DQMC results of Fig.~\ref{fig:observables2}.

Secondly, most of the EDMFT observables shown in Fig.~\ref{fig:observables:edmft} are quadratic in $V$, instead of the linear relation expected from the variational principle. This quadratic dependence has been observed and predicted before~\cite{stepanov_self-consistent_2016}. The charge susceptibility is the exception to this phenomenon, since it depends explicitly on $V$.

EDMFT has the advantage of being simpler and less demanding than DB. However, it does this at a cost. The momentum-dependence that is simplified in EDMFT is crucial for a proper description. This is especially clear when looking at the spin correlation functions.  

\section{The dynamic interaction $U(i\omega_n)$ and the effective interaction $\tilde{U}$}
\label{app:uw}

\begin{figure}
\includegraphics[]{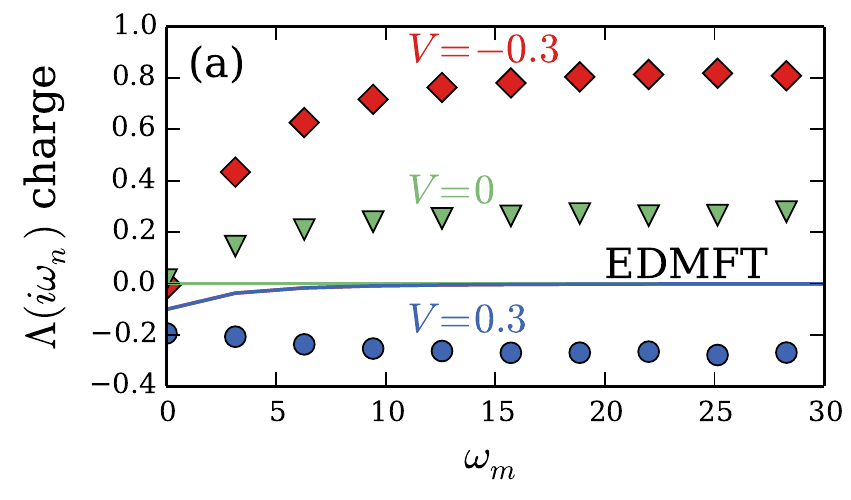}
\includegraphics[]{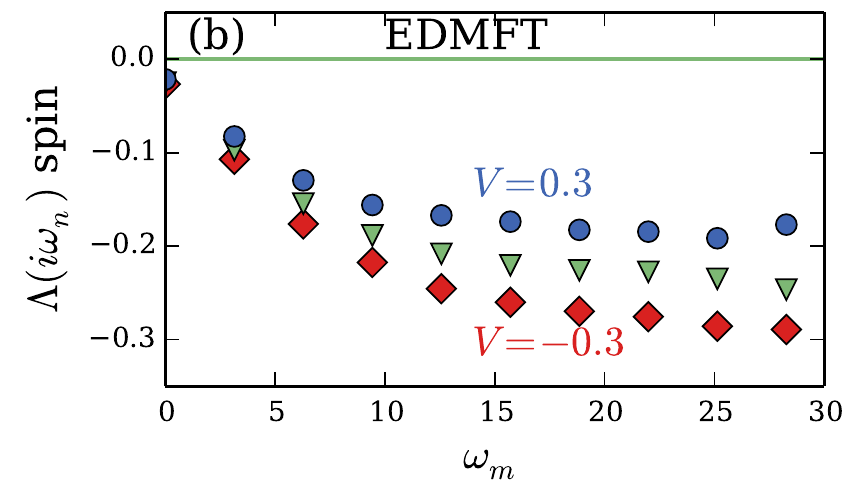}
\includegraphics[]{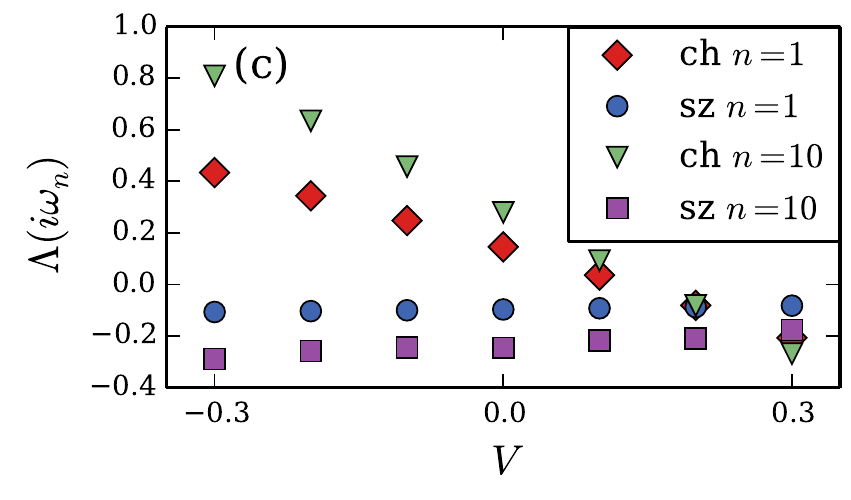}
 \caption{Dynamic interaction $\Lambda(i\omega_n)=U(i\omega_n)-U$ of the impurity model. Panels (a) and (b) show the dynamic interaction as a function of frequency, the symbols (lines) denote DB (EDMFT) results, respectively. Panel (c) shows how the dynamic interaction depends on the nearest-neighbor interaction $V$. All results are at $U=2$.}
 \label{fig:Uw}
\end{figure}

In both EDMFT and DB, the nonlocal interaction $V$ is taken into account on the level of the impurity model via a dynamic interaction. This suggests an interpretation of the renormalized, effective interaction $\tilde{U}$ in terms of the dynamic interaction $U(i\omega_n)$. In EDMFT, $U(i\omega_n=0)$ has been used~\cite{Huang14} as an estimate of $\tilde{U}$.
However, whereas the effective interaction $\tilde{U}$ is a single number, the dynamic interaction $\Lambda(i\omega_n) = U(i\omega_n) - U$ depends on the frequency and, in self-consistent DB, on the channel. In Fig.~\ref{fig:Uw}, we show $\Lambda(i\omega_n)$ in EDMFT and DB. All results are at $U=2$ and half-filling, as in Fig.~\ref{fig:observables}. 
The shape of $\Lambda(i\omega_n)$ in EDMFT has been discussed before~\cite{Ayral13,Huang14}. It is important to observe again that the EDMFT results at $V=-0.3$ and $V=+0.3$ are identical~\cite{stepanov_self-consistent_2016}. Self-consistent DB results for $\Lambda(i\omega_n)$ have appeared before, however those results were either in the charge channel only~\cite{stepanov_self-consistent_2016}, or at $V=0$ in both channels~\cite{van_loon_double_2016}. Fig.~\ref{fig:Uw}(c) shows fixed frequency results as a function of $V$. The results are all approximately linear, on the other hand the magnitude and sign of the dependence of $V$ depend on the frequency and channel. This complicates the interpretation of $\Lambda(i\omega_n)$ as the renormalized interaction strength.

If a direct comparison of the renormalized interactions in the DB (EDMFT) to the variational principle would be meaningful we should compare $\Lambda(i \omega_n)$ to $\alpha(\tilde U) V\approx 0.6 V$. Clearly, a comparison with EDMFT is not meaningful since $\Lambda(i \omega_n)\sim V^2$, to lowest order in V in EDMFT. 

The DB interaction in charge channel reveals some qualitative similarity with $\tilde U$ which is understandable from the gedankenexperiment discussed in Fig. 1 of Ref.~\onlinecite{schuler_optimal_2013}. Nonetheless, the frequency dependence of $\Lambda(i \omega_n)$ makes this comparison clearly ambiguous and the renormalization of $\Lambda(i \omega_n)$ by $V$ appears generally stronger than the change of $\tilde U$ despite the good agreement between observables calculated in the full vs. effective model shown in Fig.~\ref{fig:observables}. At the same time, the DB interaction in the spin channel shows a weaker renormalization than $\tilde U$.

Taken together, the effective interaction $\tilde U$ from the variational principle and the dynamic interaction $U(i\omega_n)$ in EDMFT (DB) are very different quantities and should not be compared directly.

\bibliographystyle{apsrev4-1}
\bibliography{newbib}

\begin{thebibliography}{42}%
\makeatletter
\providecommand \@ifxundefined [1]{%
 \@ifx{#1\undefined}
}%
\providecommand \@ifnum [1]{%
 \ifnum #1\expandafter \@firstoftwo
 \else \expandafter \@secondoftwo
 \fi
}%
\providecommand \@ifx [1]{%
 \ifx #1\expandafter \@firstoftwo
 \else \expandafter \@secondoftwo
 \fi
}%
\providecommand \natexlab [1]{#1}%
\providecommand \enquote  [1]{``#1''}%
\providecommand \bibnamefont  [1]{#1}%
\providecommand \bibfnamefont [1]{#1}%
\providecommand \citenamefont [1]{#1}%
\providecommand \href@noop [0]{\@secondoftwo}%
\providecommand \href [0]{\begingroup \@sanitize@url \@href}%
\providecommand \@href[1]{\@@startlink{#1}\@@href}%
\providecommand \@@href[1]{\endgroup#1\@@endlink}%
\providecommand \@sanitize@url [0]{\catcode `\\12\catcode `\$12\catcode
  `\&12\catcode `\#12\catcode `\^12\catcode `\_12\catcode `\%12\relax}%
\providecommand \@@startlink[1]{}%
\providecommand \@@endlink[0]{}%
\providecommand \url  [0]{\begingroup\@sanitize@url \@url }%
\providecommand \@url [1]{\endgroup\@href {#1}{\urlprefix }}%
\providecommand \urlprefix  [0]{URL }%
\providecommand \Eprint [0]{\href }%
\providecommand \doibase [0]{http://dx.doi.org/}%
\providecommand \selectlanguage [0]{\@gobble}%
\providecommand \bibinfo  [0]{\@secondoftwo}%
\providecommand \bibfield  [0]{\@secondoftwo}%
\providecommand \translation [1]{[#1]}%
\providecommand \BibitemOpen [0]{}%
\providecommand \bibitemStop [0]{}%
\providecommand \bibitemNoStop [0]{.\EOS\space}%
\providecommand \EOS [0]{\spacefactor3000\relax}%
\providecommand \BibitemShut  [1]{\csname bibitem#1\endcsname}%
\let\auto@bib@innerbib\@empty
\bibitem [{\citenamefont {Hubbard}(1963)}]{Hubbard63}%
  \BibitemOpen
  \bibfield  {author} {\bibinfo {author} {\bibfnamefont {J.}~\bibnamefont
  {Hubbard}},\ }\href {\doibase 10.1098/rspa.1963.0204} {\bibfield  {journal}
  {\bibinfo  {journal} {Proc. R. Soc. A.}\ }\textbf {\bibinfo {volume} {276}},\
  \bibinfo {pages} {238} (\bibinfo {year} {1963})}\BibitemShut {NoStop}%
\bibitem [{\citenamefont {Gutzwiller}(1963)}]{Gutzwiller63}%
  \BibitemOpen
  \bibfield  {author} {\bibinfo {author} {\bibfnamefont {M.~C.}\ \bibnamefont
  {Gutzwiller}},\ }\href {\doibase 10.1103/PhysRevLett.10.159} {\bibfield
  {journal} {\bibinfo  {journal} {Phys. Rev. Lett.}\ }\textbf {\bibinfo
  {volume} {10}},\ \bibinfo {pages} {159} (\bibinfo {year} {1963})}\BibitemShut
  {NoStop}%
\bibitem [{\citenamefont {Kanamori}(1963)}]{Kanamori63}%
  \BibitemOpen
  \bibfield  {author} {\bibinfo {author} {\bibfnamefont {J.}~\bibnamefont
  {Kanamori}},\ }\href@noop {} {\bibfield  {journal} {\bibinfo  {journal}
  {Prog. Theor. Phys.}\ }\textbf {\bibinfo {volume} {30}},\ \bibinfo {pages}
  {275} (\bibinfo {year} {1963})}\BibitemShut {NoStop}%
\bibitem [{\citenamefont {Hubbard}(1964)}]{Hubbard64}%
  \BibitemOpen
  \bibfield  {author} {\bibinfo {author} {\bibfnamefont {J.}~\bibnamefont
  {Hubbard}},\ }\href {\doibase 10.1098/rspa.1964.0190} {\bibfield  {journal}
  {\bibinfo  {journal} {Proc. R. Soc. A.}\ }\textbf {\bibinfo {volume} {281}},\
  \bibinfo {pages} {401} (\bibinfo {year} {1964})}\BibitemShut {NoStop}%
\bibitem [{\citenamefont {Gutzwiller}(1964)}]{Gutzwiller64}%
  \BibitemOpen
  \bibfield  {author} {\bibinfo {author} {\bibfnamefont {M.~C.}\ \bibnamefont
  {Gutzwiller}},\ }\href {\doibase 10.1103/PhysRev.134.A923} {\bibfield
  {journal} {\bibinfo  {journal} {Phys. Rev.}\ }\textbf {\bibinfo {volume}
  {134}},\ \bibinfo {pages} {A923} (\bibinfo {year} {1964})}\BibitemShut
  {NoStop}%
\bibitem [{\citenamefont {Peierls}(1938)}]{Peierls38}%
  \BibitemOpen
  \bibfield  {author} {\bibinfo {author} {\bibfnamefont {R.}~\bibnamefont
  {Peierls}},\ }\href {\doibase 10.1103/PhysRev.54.918} {\bibfield  {journal}
  {\bibinfo  {journal} {Phys. Rev.}\ }\textbf {\bibinfo {volume} {54}},\
  \bibinfo {pages} {918} (\bibinfo {year} {1938})}\BibitemShut {NoStop}%
\bibitem [{\citenamefont {Bogoliubov}(1958)}]{Bogoliubov58}%
  \BibitemOpen
  \bibfield  {author} {\bibinfo {author} {\bibfnamefont {N.~N.}\ \bibnamefont
  {Bogoliubov}},\ }\href@noop {} {\bibfield  {journal} {\bibinfo  {journal}
  {Dokl. Akad. Nauk SSSR}\ }\textbf {\bibinfo {volume} {119}},\ \bibinfo
  {pages} {244} (\bibinfo {year} {1958})}\BibitemShut {NoStop}%
\bibitem [{\citenamefont {Feynman}(1972)}]{Feynman72}%
  \BibitemOpen
  \bibfield  {author} {\bibinfo {author} {\bibfnamefont {R.~P.}\ \bibnamefont
  {Feynman}},\ }\href@noop {} {\emph {\bibinfo {title} {Statistical
  Mechanics}}}\ (\bibinfo  {publisher} {Benjamin, Reading Mass.},\ \bibinfo
  {year} {1972})\BibitemShut {NoStop}%
\bibitem [{\citenamefont {Sch{\"u}ler}\ \emph {et~al.}(2013)\citenamefont
  {Sch{\"u}ler}, \citenamefont {R{\"o}sner}, \citenamefont {Wehling},
  \citenamefont {Lichtenstein},\ and\ \citenamefont
  {Katsnelson}}]{schuler_optimal_2013}%
  \BibitemOpen
  \bibfield  {author} {\bibinfo {author} {\bibfnamefont {M.}~\bibnamefont
  {Sch{\"u}ler}}, \bibinfo {author} {\bibfnamefont {M.}~\bibnamefont
  {R{\"o}sner}}, \bibinfo {author} {\bibfnamefont {T.~O.}\ \bibnamefont
  {Wehling}}, \bibinfo {author} {\bibfnamefont {A.~I.}\ \bibnamefont
  {Lichtenstein}}, \ and\ \bibinfo {author} {\bibfnamefont {M.~I.}\
  \bibnamefont {Katsnelson}},\ }\href {\doibase 10.1103/PhysRevLett.111.036601}
  {\bibfield  {journal} {\bibinfo  {journal} {Physical Review Letters}\
  }\textbf {\bibinfo {volume} {111}},\ \bibinfo {pages} {036601} (\bibinfo
  {year} {2013})}\BibitemShut {NoStop}%
\bibitem [{\citenamefont {Zhang}\ and\ \citenamefont
  {Callaway}(1989)}]{zhang_extended_1989}%
  \BibitemOpen
  \bibfield  {author} {\bibinfo {author} {\bibfnamefont {Y.}~\bibnamefont
  {Zhang}}\ and\ \bibinfo {author} {\bibfnamefont {J.}~\bibnamefont
  {Callaway}},\ }\href {\doibase 10.1103/PhysRevB.39.9397} {\bibfield
  {journal} {\bibinfo  {journal} {Phys. Rev. B}\ }\textbf {\bibinfo {volume}
  {39}},\ \bibinfo {pages} {9397} (\bibinfo {year} {1989})}\BibitemShut
  {NoStop}%
\bibitem [{\citenamefont {Sun}\ and\ \citenamefont {Kotliar}(2002)}]{Sun02}%
  \BibitemOpen
  \bibfield  {author} {\bibinfo {author} {\bibfnamefont {P.}~\bibnamefont
  {Sun}}\ and\ \bibinfo {author} {\bibfnamefont {G.}~\bibnamefont {Kotliar}},\
  }\href {\doibase 10.1103/PhysRevB.66.085120} {\bibfield  {journal} {\bibinfo
  {journal} {Phys. Rev. B}\ }\textbf {\bibinfo {volume} {66}},\ \bibinfo
  {pages} {085120} (\bibinfo {year} {2002})}\BibitemShut {NoStop}%
\bibitem [{\citenamefont {Ayral}\ \emph {et~al.}(2013)\citenamefont {Ayral},
  \citenamefont {Biermann},\ and\ \citenamefont {Werner}}]{Ayral13}%
  \BibitemOpen
  \bibfield  {author} {\bibinfo {author} {\bibfnamefont {T.}~\bibnamefont
  {Ayral}}, \bibinfo {author} {\bibfnamefont {S.}~\bibnamefont {Biermann}}, \
  and\ \bibinfo {author} {\bibfnamefont {P.}~\bibnamefont {Werner}},\ }\href
  {\doibase 10.1103/PhysRevB.87.125149} {\bibfield  {journal} {\bibinfo
  {journal} {Phys. Rev. B}\ }\textbf {\bibinfo {volume} {87}},\ \bibinfo
  {pages} {125149} (\bibinfo {year} {2013})}\BibitemShut {NoStop}%
\bibitem [{\citenamefont {Huang}\ \emph {et~al.}(2014)\citenamefont {Huang},
  \citenamefont {Ayral}, \citenamefont {Biermann},\ and\ \citenamefont
  {Werner}}]{Huang14}%
  \BibitemOpen
  \bibfield  {author} {\bibinfo {author} {\bibfnamefont {L.}~\bibnamefont
  {Huang}}, \bibinfo {author} {\bibfnamefont {T.}~\bibnamefont {Ayral}},
  \bibinfo {author} {\bibfnamefont {S.}~\bibnamefont {Biermann}}, \ and\
  \bibinfo {author} {\bibfnamefont {P.}~\bibnamefont {Werner}},\ }\href
  {\doibase 10.1103/PhysRevB.90.195114} {\bibfield  {journal} {\bibinfo
  {journal} {Phys. Rev. B}\ }\textbf {\bibinfo {volume} {90}},\ \bibinfo
  {pages} {195114} (\bibinfo {year} {2014})}\BibitemShut {NoStop}%
\bibitem [{\citenamefont {van Loon}\ \emph {et~al.}(2014)\citenamefont {van
  Loon}, \citenamefont {Lichtenstein}, \citenamefont {Katsnelson},
  \citenamefont {Parcollet},\ and\ \citenamefont
  {Hafermann}}]{van_loon_beyond_2014}%
  \BibitemOpen
  \bibfield  {author} {\bibinfo {author} {\bibfnamefont {E.~G. C.~P.}\
  \bibnamefont {van Loon}}, \bibinfo {author} {\bibfnamefont {A.~I.}\
  \bibnamefont {Lichtenstein}}, \bibinfo {author} {\bibfnamefont {M.~I.}\
  \bibnamefont {Katsnelson}}, \bibinfo {author} {\bibfnamefont
  {O.}~\bibnamefont {Parcollet}}, \ and\ \bibinfo {author} {\bibfnamefont
  {H.}~\bibnamefont {Hafermann}},\ }\href {\doibase 10.1103/PhysRevB.90.235135}
  {\bibfield  {journal} {\bibinfo  {journal} {Physical Review B}\ }\textbf
  {\bibinfo {volume} {90}},\ \bibinfo {pages} {235135} (\bibinfo {year}
  {2014})}\BibitemShut {NoStop}%
\bibitem [{\citenamefont {Lhoutellier}\ \emph {et~al.}(2015)\citenamefont
  {Lhoutellier}, \citenamefont {Fr\'esard},\ and\ \citenamefont
  {Ole\ifmmode~\acute{s}\else \'{s}\fi{}}}]{Lhoutellier15}%
  \BibitemOpen
  \bibfield  {author} {\bibinfo {author} {\bibfnamefont {G.}~\bibnamefont
  {Lhoutellier}}, \bibinfo {author} {\bibfnamefont {R.}~\bibnamefont
  {Fr\'esard}}, \ and\ \bibinfo {author} {\bibfnamefont {A.~M.}\ \bibnamefont
  {Ole\ifmmode~\acute{s}\else \'{s}\fi{}}},\ }\href {\doibase
  10.1103/PhysRevB.91.224410} {\bibfield  {journal} {\bibinfo  {journal} {Phys.
  Rev. B}\ }\textbf {\bibinfo {volume} {91}},\ \bibinfo {pages} {224410}
  (\bibinfo {year} {2015})}\BibitemShut {NoStop}%
\bibitem [{\citenamefont {Frésard}\ \emph {et~al.}(2016)\citenamefont
  {Frésard}, \citenamefont {Steffen},\ and\ \citenamefont {Kopp}}]{Fresard15}%
  \BibitemOpen
  \bibfield  {author} {\bibinfo {author} {\bibfnamefont {R.}~\bibnamefont
  {Frésard}}, \bibinfo {author} {\bibfnamefont {K.}~\bibnamefont {Steffen}}, \
  and\ \bibinfo {author} {\bibfnamefont {T.}~\bibnamefont {Kopp}},\ }\href
  {http://stacks.iop.org/1742-6596/702/i=1/a=012003} {\bibfield  {journal}
  {\bibinfo  {journal} {Journal of Physics: Conference Series}\ }\textbf
  {\bibinfo {volume} {702}},\ \bibinfo {pages} {012003} (\bibinfo {year}
  {2016})}\BibitemShut {NoStop}%
\bibitem [{\citenamefont {Blankenbecler}\ \emph {et~al.}(1981)\citenamefont
  {Blankenbecler}, \citenamefont {Scalapino},\ and\ \citenamefont
  {Sugar}}]{blankenbecler_monte_1981}%
  \BibitemOpen
  \bibfield  {author} {\bibinfo {author} {\bibfnamefont {R.}~\bibnamefont
  {Blankenbecler}}, \bibinfo {author} {\bibfnamefont {D.~J.}\ \bibnamefont
  {Scalapino}}, \ and\ \bibinfo {author} {\bibfnamefont {R.~L.}\ \bibnamefont
  {Sugar}},\ }\href {\doibase 10.1103/PhysRevD.24.2278} {\bibfield  {journal}
  {\bibinfo  {journal} {Physical Review D}\ }\textbf {\bibinfo {volume} {24}},\
  \bibinfo {pages} {2278} (\bibinfo {year} {1981})}\BibitemShut {NoStop}%
\bibitem [{\citenamefont {Rubtsov}\ \emph {et~al.}(2012)\citenamefont
  {Rubtsov}, \citenamefont {Katsnelson},\ and\ \citenamefont
  {Lichtenstein}}]{rubtsov_dual_2012}%
  \BibitemOpen
  \bibfield  {author} {\bibinfo {author} {\bibfnamefont {A.~N.}\ \bibnamefont
  {Rubtsov}}, \bibinfo {author} {\bibfnamefont {M.~I.}\ \bibnamefont
  {Katsnelson}}, \ and\ \bibinfo {author} {\bibfnamefont {A.~I.}\ \bibnamefont
  {Lichtenstein}},\ }\href {\doibase 10.1016/j.aop.2012.01.002} {\bibfield
  {journal} {\bibinfo  {journal} {Annals of Physics}\ }\textbf {\bibinfo
  {volume} {327}},\ \bibinfo {pages} {1320} (\bibinfo {year}
  {2012})}\BibitemShut {NoStop}%
\bibitem [{\citenamefont {Stepanov}\ \emph {et~al.}(2016)\citenamefont
  {Stepanov}, \citenamefont {van Loon}, \citenamefont {Katanin}, \citenamefont
  {Lichtenstein}, \citenamefont {Katsnelson},\ and\ \citenamefont
  {Rubtsov}}]{stepanov_self-consistent_2016}%
  \BibitemOpen
  \bibfield  {author} {\bibinfo {author} {\bibfnamefont {E.~A.}\ \bibnamefont
  {Stepanov}}, \bibinfo {author} {\bibfnamefont {E.~G. C.~P.}\ \bibnamefont
  {van Loon}}, \bibinfo {author} {\bibfnamefont {A.~A.}\ \bibnamefont
  {Katanin}}, \bibinfo {author} {\bibfnamefont {A.~I.}\ \bibnamefont
  {Lichtenstein}}, \bibinfo {author} {\bibfnamefont {M.~I.}\ \bibnamefont
  {Katsnelson}}, \ and\ \bibinfo {author} {\bibfnamefont {A.~N.}\ \bibnamefont
  {Rubtsov}},\ }\href {\doibase 10.1103/PhysRevB.93.045107} {\bibfield
  {journal} {\bibinfo  {journal} {Phys. Rev. B}\ }\textbf {\bibinfo {volume}
  {93}},\ \bibinfo {pages} {045107} (\bibinfo {year} {2016})}\BibitemShut
  {NoStop}%
\bibitem [{\citenamefont {Metzner}\ and\ \citenamefont
  {Vollhardt}(1989)}]{metzner89}%
  \BibitemOpen
  \bibfield  {author} {\bibinfo {author} {\bibfnamefont {W.}~\bibnamefont
  {Metzner}}\ and\ \bibinfo {author} {\bibfnamefont {D.}~\bibnamefont
  {Vollhardt}},\ }\href {\doibase 10.1103/PhysRevLett.62.324} {\bibfield
  {journal} {\bibinfo  {journal} {Phys. Rev. Lett.}\ }\textbf {\bibinfo
  {volume} {62}},\ \bibinfo {pages} {324} (\bibinfo {year} {1989})}\BibitemShut
  {NoStop}%
\bibitem [{\citenamefont {Georges}\ \emph {et~al.}(1996)\citenamefont
  {Georges}, \citenamefont {Kotliar}, \citenamefont {Krauth},\ and\
  \citenamefont {Rozenberg}}]{georges_dynamical_1996}%
  \BibitemOpen
  \bibfield  {author} {\bibinfo {author} {\bibfnamefont {A.}~\bibnamefont
  {Georges}}, \bibinfo {author} {\bibfnamefont {G.}~\bibnamefont {Kotliar}},
  \bibinfo {author} {\bibfnamefont {W.}~\bibnamefont {Krauth}}, \ and\ \bibinfo
  {author} {\bibfnamefont {M.~J.}\ \bibnamefont {Rozenberg}},\ }\href {\doibase
  10.1103/RevModPhys.68.13} {\bibfield  {journal} {\bibinfo  {journal} {Rev.
  Mod. Phys.}\ }\textbf {\bibinfo {volume} {68}},\ \bibinfo {pages} {13}
  (\bibinfo {year} {1996})}\BibitemShut {NoStop}%
\bibitem [{\citenamefont {Mahan}(2000)}]{mahan00}%
  \BibitemOpen
  \bibfield  {author} {\bibinfo {author} {\bibfnamefont {G.~D.}\ \bibnamefont
  {Mahan}},\ }\href@noop {} {\emph {\bibinfo {title} {Many-Particle
  Physics}}},\ Physics of Solids and Liquids\ (\bibinfo  {publisher} {Plenum},\
  \bibinfo {year} {2000})\ \bibinfo {note} {3rd ed.}\BibitemShut {Stop}%
\bibitem [{\citenamefont {{Schubin}}\ and\ \citenamefont
  {{Wonsowsky}}(1934)}]{Schubin34}%
  \BibitemOpen
  \bibfield  {author} {\bibinfo {author} {\bibfnamefont {S.}~\bibnamefont
  {{Schubin}}}\ and\ \bibinfo {author} {\bibfnamefont {S.}~\bibnamefont
  {{Wonsowsky}}},\ }\href {\doibase 10.1098/rspa.1934.0089} {\bibfield
  {journal} {\bibinfo  {journal} {Proceedings of the Royal Society of London
  Series A}\ }\textbf {\bibinfo {volume} {145}},\ \bibinfo {pages} {159}
  (\bibinfo {year} {1934})}\BibitemShut {NoStop}%
\bibitem [{\citenamefont {Vonsovsky}\ and\ \citenamefont
  {Katsnelson}(1979{\natexlab{a}})}]{Vonsovsky79_1}%
  \BibitemOpen
  \bibfield  {author} {\bibinfo {author} {\bibfnamefont {S.~V.}\ \bibnamefont
  {Vonsovsky}}\ and\ \bibinfo {author} {\bibfnamefont {M.~I.}\ \bibnamefont
  {Katsnelson}},\ }\href {http://stacks.iop.org/0022-3719/12/i=11/a=015}
  {\bibfield  {journal} {\bibinfo  {journal} {Journal of Physics C: Solid State
  Physics}\ }\textbf {\bibinfo {volume} {12}},\ \bibinfo {pages} {2043}
  (\bibinfo {year} {1979}{\natexlab{a}})}\BibitemShut {NoStop}%
\bibitem [{\citenamefont {Vonsovsky}\ and\ \citenamefont
  {Katsnelson}(1979{\natexlab{b}})}]{Vonsovsky79_2}%
  \BibitemOpen
  \bibfield  {author} {\bibinfo {author} {\bibfnamefont {S.~V.}\ \bibnamefont
  {Vonsovsky}}\ and\ \bibinfo {author} {\bibfnamefont {M.~I.}\ \bibnamefont
  {Katsnelson}},\ }\href {http://stacks.iop.org/0022-3719/12/i=11/a=016}
  {\bibfield  {journal} {\bibinfo  {journal} {Journal of Physics C: Solid State
  Physics}\ }\textbf {\bibinfo {volume} {12}},\ \bibinfo {pages} {2055}
  (\bibinfo {year} {1979}{\natexlab{b}})}\BibitemShut {NoStop}%
\bibitem [{Note1()}]{Note1}%
  \BibitemOpen
  \bibinfo {note} {``QUantum Electron Simulation Toolbox'' \protect \textsc
  {quest} 1.3.0 A. Tomas, C-C. Chang, Z-J. Bai, and R. Scalettar, (\protect
  \url {http://quest.ucdavis.edu/})}\BibitemShut {NoStop}%
\bibitem [{\citenamefont {Dopf}\ \emph {et~al.}(1992)\citenamefont {Dopf},
  \citenamefont {Muramatsu},\ and\ \citenamefont {Hanke}}]{PhysRevLett.68.353}%
  \BibitemOpen
  \bibfield  {author} {\bibinfo {author} {\bibfnamefont {G.}~\bibnamefont
  {Dopf}}, \bibinfo {author} {\bibfnamefont {A.}~\bibnamefont {Muramatsu}}, \
  and\ \bibinfo {author} {\bibfnamefont {W.}~\bibnamefont {Hanke}},\ }\href
  {\doibase 10.1103/PhysRevLett.68.353} {\bibfield  {journal} {\bibinfo
  {journal} {Phys. Rev. Lett.}\ }\textbf {\bibinfo {volume} {68}},\ \bibinfo
  {pages} {353} (\bibinfo {year} {1992})}\BibitemShut {NoStop}%
\bibitem [{\citenamefont {Golor}\ and\ \citenamefont
  {Wessel}(2015)}]{golor_nonlocal_2015}%
  \BibitemOpen
  \bibfield  {author} {\bibinfo {author} {\bibfnamefont {M.}~\bibnamefont
  {Golor}}\ and\ \bibinfo {author} {\bibfnamefont {S.}~\bibnamefont {Wessel}},\
  }\href {\doibase 10.1103/PhysRevB.92.195154} {\bibfield  {journal} {\bibinfo
  {journal} {Physical Review B}\ }\textbf {\bibinfo {volume} {92}},\ \bibinfo
  {pages} {195154} (\bibinfo {year} {2015})}\BibitemShut {NoStop}%
\bibitem [{\citenamefont {Pavarini}\ \emph {et~al.}(2014)\citenamefont
  {Pavarini}, \citenamefont {Koch}, \citenamefont {Vollhardt},\ and\
  \citenamefont {Lichtenstein}}]{PavariniJulich}%
  \BibitemOpen
  \bibinfo {editor} {\bibfnamefont {E.}~\bibnamefont {Pavarini}}, \bibinfo
  {editor} {\bibfnamefont {E.}~\bibnamefont {Koch}}, \bibinfo {editor}
  {\bibfnamefont {D.}~\bibnamefont {Vollhardt}}, \ and\ \bibinfo {editor}
  {\bibfnamefont {A.}~\bibnamefont {Lichtenstein}},\ eds.,\ \href
  {http://juser.fz-juelich.de/record/155829} {\emph {\bibinfo {title} {{DMFT}
  at 25: {I}nfinite {D}imensions}}},\ \bibinfo {series} {Modeling and
  Simulation}, Vol.~\bibinfo {volume} {4},\ \bibinfo {organization} {Autumn
  School on Correlated Electrons, Jülich (Germany), 15 Sep 2014 - 19 Sep
  2014}\ (\bibinfo  {publisher} {Forschungszentrum Jülich Zentralbibliothek,
  Verlag},\ \bibinfo {address} {Jülich},\ \bibinfo {year} {2014})\BibitemShut
  {NoStop}%
\bibitem [{\citenamefont {van Loon}\ \emph {et~al.}(2016)\citenamefont {van
  Loon}, \citenamefont {Krien}, \citenamefont {Hafermann}, \citenamefont
  {Stepanov}, \citenamefont {Lichtenstein},\ and\ \citenamefont
  {Katsnelson}}]{van_loon_double_2016}%
  \BibitemOpen
  \bibfield  {author} {\bibinfo {author} {\bibfnamefont {E.~G. C.~P.}\
  \bibnamefont {van Loon}}, \bibinfo {author} {\bibfnamefont {F.}~\bibnamefont
  {Krien}}, \bibinfo {author} {\bibfnamefont {H.}~\bibnamefont {Hafermann}},
  \bibinfo {author} {\bibfnamefont {E.~A.}\ \bibnamefont {Stepanov}}, \bibinfo
  {author} {\bibfnamefont {A.~I.}\ \bibnamefont {Lichtenstein}}, \ and\
  \bibinfo {author} {\bibfnamefont {M.~I.}\ \bibnamefont {Katsnelson}},\ }\href
  {\doibase 10.1103/PhysRevB.93.155162} {\bibfield  {journal} {\bibinfo
  {journal} {Phys. Rev. B}\ }\textbf {\bibinfo {volume} {93}},\ \bibinfo
  {pages} {155162} (\bibinfo {year} {2016})}\BibitemShut {NoStop}%
\bibitem [{\citenamefont {Werner}\ \emph {et~al.}(2006)\citenamefont {Werner},
  \citenamefont {Comanac}, \citenamefont {de' Medici}, \citenamefont {Troyer},\
  and\ \citenamefont {Millis}}]{Werner06}%
  \BibitemOpen
  \bibfield  {author} {\bibinfo {author} {\bibfnamefont {P.}~\bibnamefont
  {Werner}}, \bibinfo {author} {\bibfnamefont {A.}~\bibnamefont {Comanac}},
  \bibinfo {author} {\bibfnamefont {L.}~\bibnamefont {de' Medici}}, \bibinfo
  {author} {\bibfnamefont {M.}~\bibnamefont {Troyer}}, \ and\ \bibinfo {author}
  {\bibfnamefont {A.~J.}\ \bibnamefont {Millis}},\ }\href {\doibase
  10.1103/PhysRevLett.97.076405} {\bibfield  {journal} {\bibinfo  {journal}
  {Phys. Rev. Lett.}\ }\textbf {\bibinfo {volume} {97}},\ \bibinfo {pages}
  {076405} (\bibinfo {year} {2006})}\BibitemShut {NoStop}%
\bibitem [{\citenamefont {Bauer}\ \emph {et~al.}(2011)\citenamefont {Bauer},
  \citenamefont {Carr}, \citenamefont {Evertz}, \citenamefont {Feiguin},
  \citenamefont {Freire}, \citenamefont {Fuchs}, \citenamefont {Gamper},
  \citenamefont {Gukelberger}, \citenamefont {Gull}, \citenamefont {Guertler},
  \citenamefont {Hehn}, \citenamefont {Igarashi}, \citenamefont {Isakov},
  \citenamefont {Koop}, \citenamefont {Ma}, \citenamefont {Mates},
  \citenamefont {Matsuo}, \citenamefont {Parcollet}, \citenamefont
  {Pawłowski}, \citenamefont {Picon}, \citenamefont {Pollet}, \citenamefont
  {Santos}, \citenamefont {Scarola}, \citenamefont {Schollwöck}, \citenamefont
  {Silva}, \citenamefont {Surer}, \citenamefont {Todo}, \citenamefont {Trebst},
  \citenamefont {Troyer}, \citenamefont {Wall}, \citenamefont {Werner},\ and\
  \citenamefont {Wessel}}]{ALPS2}%
  \BibitemOpen
  \bibfield  {author} {\bibinfo {author} {\bibfnamefont {B.}~\bibnamefont
  {Bauer}}, \bibinfo {author} {\bibfnamefont {L.~D.}\ \bibnamefont {Carr}},
  \bibinfo {author} {\bibfnamefont {H.~G.}\ \bibnamefont {Evertz}}, \bibinfo
  {author} {\bibfnamefont {A.}~\bibnamefont {Feiguin}}, \bibinfo {author}
  {\bibfnamefont {J.}~\bibnamefont {Freire}}, \bibinfo {author} {\bibfnamefont
  {S.}~\bibnamefont {Fuchs}}, \bibinfo {author} {\bibfnamefont
  {L.}~\bibnamefont {Gamper}}, \bibinfo {author} {\bibfnamefont
  {J.}~\bibnamefont {Gukelberger}}, \bibinfo {author} {\bibfnamefont
  {E.}~\bibnamefont {Gull}}, \bibinfo {author} {\bibfnamefont {S.}~\bibnamefont
  {Guertler}}, \bibinfo {author} {\bibfnamefont {A.}~\bibnamefont {Hehn}},
  \bibinfo {author} {\bibfnamefont {R.}~\bibnamefont {Igarashi}}, \bibinfo
  {author} {\bibfnamefont {S.~V.}\ \bibnamefont {Isakov}}, \bibinfo {author}
  {\bibfnamefont {D.}~\bibnamefont {Koop}}, \bibinfo {author} {\bibfnamefont
  {P.~N.}\ \bibnamefont {Ma}}, \bibinfo {author} {\bibfnamefont
  {P.}~\bibnamefont {Mates}}, \bibinfo {author} {\bibfnamefont
  {H.}~\bibnamefont {Matsuo}}, \bibinfo {author} {\bibfnamefont
  {O.}~\bibnamefont {Parcollet}}, \bibinfo {author} {\bibfnamefont
  {G.}~\bibnamefont {Pawłowski}}, \bibinfo {author} {\bibfnamefont {J.~D.}\
  \bibnamefont {Picon}}, \bibinfo {author} {\bibfnamefont {L.}~\bibnamefont
  {Pollet}}, \bibinfo {author} {\bibfnamefont {E.}~\bibnamefont {Santos}},
  \bibinfo {author} {\bibfnamefont {V.~W.}\ \bibnamefont {Scarola}}, \bibinfo
  {author} {\bibfnamefont {U.}~\bibnamefont {Schollwöck}}, \bibinfo {author}
  {\bibfnamefont {C.}~\bibnamefont {Silva}}, \bibinfo {author} {\bibfnamefont
  {B.}~\bibnamefont {Surer}}, \bibinfo {author} {\bibfnamefont
  {S.}~\bibnamefont {Todo}}, \bibinfo {author} {\bibfnamefont {S.}~\bibnamefont
  {Trebst}}, \bibinfo {author} {\bibfnamefont {M.}~\bibnamefont {Troyer}},
  \bibinfo {author} {\bibfnamefont {M.~L.}\ \bibnamefont {Wall}}, \bibinfo
  {author} {\bibfnamefont {P.}~\bibnamefont {Werner}}, \ and\ \bibinfo {author}
  {\bibfnamefont {S.}~\bibnamefont {Wessel}},\ }\href
  {http://stacks.iop.org/1742-5468/2011/i=05/a=P05001} {\bibfield  {journal}
  {\bibinfo  {journal} {Journal of Statistical Mechanics: Theory and
  Experiment}\ }\textbf {\bibinfo {volume} {2011}},\ \bibinfo {pages} {P05001}
  (\bibinfo {year} {2011})}\BibitemShut {NoStop}%
\bibitem [{\citenamefont {Hafermann}\ \emph {et~al.}(2013)\citenamefont
  {Hafermann}, \citenamefont {Werner},\ and\ \citenamefont
  {Gull}}]{Hafermann13}%
  \BibitemOpen
  \bibfield  {author} {\bibinfo {author} {\bibfnamefont {H.}~\bibnamefont
  {Hafermann}}, \bibinfo {author} {\bibfnamefont {P.}~\bibnamefont {Werner}}, \
  and\ \bibinfo {author} {\bibfnamefont {E.}~\bibnamefont {Gull}},\ }\href
  {\doibase http://dx.doi.org/10.1016/j.cpc.2012.12.013} {\bibfield  {journal}
  {\bibinfo  {journal} {Computer Physics Communications}\ }\textbf {\bibinfo
  {volume} {184}},\ \bibinfo {pages} {1280 } (\bibinfo {year}
  {2013})}\BibitemShut {NoStop}%
\bibitem [{\citenamefont {Hafermann}(2014)}]{Hafermann14}%
  \BibitemOpen
  \bibfield  {author} {\bibinfo {author} {\bibfnamefont {H.}~\bibnamefont
  {Hafermann}},\ }\href {\doibase 10.1103/PhysRevB.89.235128} {\bibfield
  {journal} {\bibinfo  {journal} {Phys. Rev. B}\ }\textbf {\bibinfo {volume}
  {89}},\ \bibinfo {pages} {235128} (\bibinfo {year} {2014})}\BibitemShut
  {NoStop}%
\bibitem [{\citenamefont {Sch\"afer}\ \emph {et~al.}(2015)\citenamefont
  {Sch\"afer}, \citenamefont {Geles}, \citenamefont {Rost}, \citenamefont
  {Rohringer}, \citenamefont {Arrigoni}, \citenamefont {Held}, \citenamefont
  {Bl\"umer}, \citenamefont {Aichhorn},\ and\ \citenamefont
  {Toschi}}]{Schafer15}%
  \BibitemOpen
  \bibfield  {author} {\bibinfo {author} {\bibfnamefont {T.}~\bibnamefont
  {Sch\"afer}}, \bibinfo {author} {\bibfnamefont {F.}~\bibnamefont {Geles}},
  \bibinfo {author} {\bibfnamefont {D.}~\bibnamefont {Rost}}, \bibinfo {author}
  {\bibfnamefont {G.}~\bibnamefont {Rohringer}}, \bibinfo {author}
  {\bibfnamefont {E.}~\bibnamefont {Arrigoni}}, \bibinfo {author}
  {\bibfnamefont {K.}~\bibnamefont {Held}}, \bibinfo {author} {\bibfnamefont
  {N.}~\bibnamefont {Bl\"umer}}, \bibinfo {author} {\bibfnamefont
  {M.}~\bibnamefont {Aichhorn}}, \ and\ \bibinfo {author} {\bibfnamefont
  {A.}~\bibnamefont {Toschi}},\ }\href {\doibase 10.1103/PhysRevB.91.125109}
  {\bibfield  {journal} {\bibinfo  {journal} {Phys. Rev. B}\ }\textbf {\bibinfo
  {volume} {91}},\ \bibinfo {pages} {125109} (\bibinfo {year}
  {2015})}\BibitemShut {NoStop}%
\bibitem [{\citenamefont {Wigner}(1934)}]{wigner34}%
  \BibitemOpen
  \bibfield  {author} {\bibinfo {author} {\bibfnamefont {E.}~\bibnamefont
  {Wigner}},\ }\href {\doibase 10.1103/PhysRev.46.1002} {\bibfield  {journal}
  {\bibinfo  {journal} {Phys. Rev.}\ }\textbf {\bibinfo {volume} {46}},\
  \bibinfo {pages} {1002} (\bibinfo {year} {1934})}\BibitemShut {NoStop}%
\bibitem [{\citenamefont {van Loon}\ \emph {et~al.}(2015)\citenamefont {van
  Loon}, \citenamefont {Hafermann}, \citenamefont {Lichtenstein},\ and\
  \citenamefont {Katsnelson}}]{van_loon_thermodynamic_2015}%
  \BibitemOpen
  \bibfield  {author} {\bibinfo {author} {\bibfnamefont {E.~G. C.~P.}\
  \bibnamefont {van Loon}}, \bibinfo {author} {\bibfnamefont {H.}~\bibnamefont
  {Hafermann}}, \bibinfo {author} {\bibfnamefont {A.~I.}\ \bibnamefont
  {Lichtenstein}}, \ and\ \bibinfo {author} {\bibfnamefont {M.~I.}\
  \bibnamefont {Katsnelson}},\ }\href {\doibase 10.1103/PhysRevB.92.085106}
  {\bibfield  {journal} {\bibinfo  {journal} {Phys. Rev. B}\ }\textbf {\bibinfo
  {volume} {92}},\ \bibinfo {pages} {085106} (\bibinfo {year}
  {2015})}\BibitemShut {NoStop}%
\bibitem [{\citenamefont {Si}\ and\ \citenamefont {Smith}(1996)}]{Si96}%
  \BibitemOpen
  \bibfield  {author} {\bibinfo {author} {\bibfnamefont {Q.}~\bibnamefont
  {Si}}\ and\ \bibinfo {author} {\bibfnamefont {J.~L.}\ \bibnamefont {Smith}},\
  }\href {\doibase 10.1103/PhysRevLett.77.3391} {\bibfield  {journal} {\bibinfo
   {journal} {Phys. Rev. Lett.}\ }\textbf {\bibinfo {volume} {77}},\ \bibinfo
  {pages} {3391} (\bibinfo {year} {1996})}\BibitemShut {NoStop}%
\bibitem [{\citenamefont {Kajueter}(1996)}]{Kajueter96}%
  \BibitemOpen
  \bibfield  {author} {\bibinfo {author} {\bibfnamefont {H.}~\bibnamefont
  {Kajueter}},\ }\href@noop {} {Ph.D. thesis},\ \bibinfo  {school} {Rutgers
  University} (\bibinfo {year} {1996})\BibitemShut {NoStop}%
\bibitem [{\citenamefont {Smith}\ and\ \citenamefont {Si}(2000)}]{Smith00}%
  \BibitemOpen
  \bibfield  {author} {\bibinfo {author} {\bibfnamefont {J.~L.}\ \bibnamefont
  {Smith}}\ and\ \bibinfo {author} {\bibfnamefont {Q.}~\bibnamefont {Si}},\
  }\href {\doibase 10.1103/PhysRevB.61.5184} {\bibfield  {journal} {\bibinfo
  {journal} {Phys. Rev. B}\ }\textbf {\bibinfo {volume} {61}},\ \bibinfo
  {pages} {5184} (\bibinfo {year} {2000})}\BibitemShut {NoStop}%
\bibitem [{\citenamefont {Chitra}\ and\ \citenamefont
  {Kotliar}(2000)}]{Chitra00}%
  \BibitemOpen
  \bibfield  {author} {\bibinfo {author} {\bibfnamefont {R.}~\bibnamefont
  {Chitra}}\ and\ \bibinfo {author} {\bibfnamefont {G.}~\bibnamefont
  {Kotliar}},\ }\href {\doibase 10.1103/PhysRevLett.84.3678} {\bibfield
  {journal} {\bibinfo  {journal} {Phys. Rev. Lett.}\ }\textbf {\bibinfo
  {volume} {84}},\ \bibinfo {pages} {3678} (\bibinfo {year}
  {2000})}\BibitemShut {NoStop}%
\bibitem [{\citenamefont {Chitra}\ and\ \citenamefont
  {Kotliar}(2001)}]{Chitra01}%
  \BibitemOpen
  \bibfield  {author} {\bibinfo {author} {\bibfnamefont {R.}~\bibnamefont
  {Chitra}}\ and\ \bibinfo {author} {\bibfnamefont {G.}~\bibnamefont
  {Kotliar}},\ }\href {\doibase 10.1103/PhysRevB.63.115110} {\bibfield
  {journal} {\bibinfo  {journal} {Phys. Rev. B}\ }\textbf {\bibinfo {volume}
  {63}},\ \bibinfo {pages} {115110} (\bibinfo {year} {2001})}\BibitemShut
  {NoStop}%
\end{thebibliography}%

\end{document}